\documentclass[aps,twocolumn,superscriptaddress,floatfix,prx,10pt]{revtex4-2}

\usepackage{graphicx}

\usepackage[utf8]{inputenc}
\usepackage{amssymb}
\usepackage{amsmath}
\usepackage[colorlinks,allcolors=blue]{hyperref}
\usepackage{braket}
\usepackage{placeins}

\usepackage{soul}

\begin{document}

\setlength\abovedisplayskip{5pt}
\setlength\belowdisplayskip{5pt}

\title{Quantum-classical correspondence of strongly chaotic many-body spin models}
\author{Luis Benet}
\affiliation{Instituto de Ciencias F\'{i}sicas, Universidad Nacional Aut\'{o}noma de M\'{e}xico (UNAM), Av. Universidad s/n, Col. Chamilpa, CP 62210 Cuernavaca, Mor., Mexico}
\author{Fausto Borgonovi}
\affiliation{Dipartimento di Matematica e
  Fisica and Interdisciplinary Laboratories for Advanced Materials Physics,
  Universit\`a Cattolica, via della Garzetta  48, 25133 Brescia, Italy}
\affiliation{Istituto Nazionale di Fisica Nucleare,  Sezione di Milano,
  via Celoria 16, I-20133,  Milano, Italy}
\author{Felix M. Izrailev}
\affiliation{Instituto de F\'{i}sica, Benem\'{e}rita Universidad Aut\'{o}noma
  de Puebla, Apartado Postal J-48, Puebla 72570, Mexico}
\affiliation{Dept. of Physics and Astronomy, Michigan State University, E. Lansing, Michigan 48824-1321, USA}
\author{Lea F. Santos}
\affiliation{Department of Physics, University of Connecticut, Storrs, CT, USA}

\date{\today}

\begin{abstract}
We study the quantum-classical correspondence for systems with interacting spin-particles that are strongly chaotic in the classical limit. This is done in the presence of constants of motion associated with the fixed angular momenta of individual spins. Our analysis of the Lyapunov spectra  reveals that the largest Lyapunov exponent agrees with the Lyapunov exponent that determines the local instability of each individual spin moving under the influence of all other spins. Within this picture, we introduce a rigorous and simple test of ergodicity for the spin motion, and use it  to identify when classical chaos is both strong and global in phase space.  In the quantum domain, our analysis of the Hamiltonian matrix in a proper representation allows us to obtain the conditions for the onset of quantum chaos as a function of the model parameters. From the comparison between the quantum and classical domains, we demonstrate that quantum quantities, such as the local density of states (LDOS) and the shape of the chaotic eigenfunctions written in the non-interacting many-body basis, have well-defined classical counterparts. Another central finding is the relationship between the Kolmogorov-Sinai entropy and the width of the LDOS, which is useful for studies of many-body dynamics.
\end{abstract}

\maketitle

%%%%%%%%% INTRO %%%%%%%%%%%%
\section{Introduction}

The quantum-classical correspondence (QCC) principle is one of the deepest concepts of physics that has attracted significant attention since the birth of quantum physics. The different nature between classical and quantum mechanics is at the origin of the QCC problem; this difference being highly pronounced in deterministic Hamiltonian systems.  One of the first rigorous results connecting the quantum and classical domains and often mentioned in the current literature is the Ehrenfest theorem~\cite{Ehrenfest1927}. It states that narrow quantum packets propagate along the classical trajectories for a finite timescale during which the spread of the packets in the phase space can be neglected. 

Until the birth of the theory of quantum chaos, the Ehrenfest theorem served as the main tool for establishing the QCC. However, the application of this theorem to quantum systems that exhibit chaotic motion in the classical limit led to unexpected results. As analytically shown in~\cite{Berman1978}, when the system is classically chaotic, the quantum packets spread exponentially fast resulting in a very short physical timescale $t_E$ on which there is a complete correspondence between the quantum and classical descriptions of the  deterministic systems' behaviors (see also discussion in \cite{Chirikov1981}). 

The estimate of the Ehrenfest time for the simplest one-dimensional chaotic systems is given by the expression $t_E\propto \lambda^{-1}\ln({1/\hbar_{\text{eff}}})$, which depends on {\it both} the classical Lyapunov exponent $\lambda$ and the effective quantum parameter $\hbar_{\text{eff}}$ proportional to the Planck constant $\hbar$. This expression reveals two cornerstones in the theory of the QCC. The first one is that the more unstable the classical motion is, the shorter the time $t_E$ becomes. The second is that by moving deeper into the quantum region, the Ehrenfest time $t_E$ shrinks. 

The first numerical study of a paradigmatic model of classical and quantum chaos,  the kicked rotor~\cite{Casati1979}, revealed another timescale, $t_D\gg t_E$, on which there is a good QCC for the width of the quantum packet as it diffusively spreads in momentum space. Specifically, it was found that the process of quantum diffusion gradually slows down, and for times $t \gg t_D$, a complete saturation takes place, while classical diffusion continues. The semi-analytical approach developed in \cite{Chirikov1981,Chirikov1988} showed that the saturation results from the localization of the eigenstates that are involved in the dynamics. Due to the finite size of the eigenstates in the infinite momentum space, only a fraction of the phase space can be covered by the quantum packet. Measuring the localization length $l_{\infty}$ of the eigenstates by projecting them onto the unperturbed basis in which the dynamics is explored, it was found numerically and explained semi-analytically that  $l_{\infty}$ is proportional to the classical diffusion coefficient $D_{cl}$. This direct link between classical diffusion and the localization of the eigenstates is a generic property of quantum chaos that became known as dynamical localization. It is observed in various disordered models and is a main finding in the theory of quantum chaos.

As a result of the discovery of the two timescales on which the chaotic properties of motion get manifested in a quantum system, $t_E$ and $t_D$, the question arose about when the correspondence principle holds. Local properties of chaos are evident in quantum systems on timescales $t\lesssim t_E$, while for global properties, this happens for $t < t_D$. This means that in quantum systems, manifestations of chaos depend on the time at which the evolution of the system is considered. Thus, we have to agree that there is no unique answer to the validity of the correspondence principle itself. The answer does not depend only on the generally accepted classical limit $\hbar_{\text{eff}} \rightarrow 0$,  but also on how the limit is taken, whether before or after the time limit $t \rightarrow \infty$. If, for example, we fix $\hbar_{\text{eff}}$ and let the time go to infinity first,  the correspondence principle will be violated. If, however, we assume that the correct procedure is to take $\hbar_{\text{eff}} \rightarrow 0$ first, we then fix a finite time at which the properties of the system are considered, and take $t \rightarrow \infty$ only after taking $\hbar_{\text{eff}} \rightarrow 0$, therefore ``saving'' the correspondence principle as applied to chaotic systems. (The question of which limit should be taken first arises also in classical physics when investigating integrable nonlinear systems with a large number $N$ of degrees of freedom, the issue in this case being the order of the limits $N \rightarrow \infty$ and  $t \rightarrow \infty$.)

Taking $\hbar_{\text{eff}} \rightarrow 0$ first was the solution to the QCC problem proposed by Chirikov at the beginning of the quantum chaos theory~\cite{Chirikov1981}. According to Chirikov, chaotic properties are observed in quantum systems at finite times only, be the time short or long. Similar points have been partially addressed in several recent studies~\cite{Levstein1998,Cucchietti2002,Wisniacki2003, Cucchietti2004,Gorin2006,Wijn2012,Elsayed2015,Rozenbaum2017,GarciaMata2018,Pappalardi2018,Chavez2019,Rammensee2018,Hummel2019,Lakshminarayan2019,Pilatowsky2020,Wang2020,Xu2020,Hashimoto2020,Wang2021} that relate  unstable classical motion  with the exponentially fast evolution of quantum observables, including the out-of-time ordered correlator (OTOC) \cite{Larkin1969}. In these works, the QCC is mostly restricted to short times, typically on the same order as the Ehrenfest time.

Further numerical studies comparing classical and quantum diffusion~\cite{Shepelyansky1983} led to the conclusion that the observed quantum diffusion, although  nicely reproducing the global properties of classical diffusion, is not {\it true} diffusion, because quantum diffusion is reversible in time, in contrast to diffusion in classical physics. By numerically reversing the evolution, the initial quantum packet is completely recovered, while this is not possible in a classical system due to the exponential sensitivity of the chaotic motion to any weak perturbation. In essence, chaos that occurs in quantum systems at times $t > t_E$ for some global observables has a different nature from chaos emerging in classical physics. 

Since the time evolution of quantum systems is entirely determined by the properties of the energy (or quasi-energy) spectrum and the eigenstates, many efforts have been made to relate essential properties of a (pseudo)-chaotic quantum dynamics to the statistical properties of spectra and eigenstates. One of the earliest suggestions in this direction was Berry's conjecture that for quantum billiards fully chaotic in the classical limit, the eigenstates may be treated as random superpositions of plane waves \cite{Berry1977}. Therefore, quantum chaos can be understood as the  complicated (random) structure of the stationary states in a physically chosen basis (see also~\cite{Flambaum1994,Zelevinsky1995,Horoi1995,ZelevinskyRep1996,Flambaum2001b,Borgonovi2016}). 
As for the eigenvalues, the properties of the energy spectra of quantum systems that are strongly chaotic in the classical limit should be comparable to those determined by random matrices of a specific symmetry \cite{Casati1980,Bohigas1984}. 

Originally, the interest in the statistical properties of the energy spectra of complex quantum systems emerged before the birth of quantum chaos. It was motivated by experimental studies of heavy nuclei and many-electron atoms (see, e.g., \cite{Guhr1998,Firk2009} and references therein). One of the first questions addressed was the shape of the distribution $P(s)$ of the spacings $s$ between nearest energy levels in relation to the results from experiments with nucleon scattering on nuclei. It was observed that the probability of small values of $s$ decreases with $s$, thus manifesting repulsion between nearest neighboring energy levels. After intensive discussions about the form of $P(s)$, Wigner suggested the expression that is nowadays known as the Wigner surmise~\cite{Wigner1951p}, obtained with the use of simple scaling arguments. Later, he indicated that the form of $P(s)$ could be explained within the theory of random matrices. Further studies of random matrices (see the collection of papers in \cite{PorterBook}) showed that the degree of repulsion for small spacings $s$ depends on the symmetry of the random matrices or, equivalently, on the underlying symmetry of the physical systems reflected by the symmetry of their Hamiltonians. Although there is no analytical expression valid for any value of $s$, it was numerically shown~\cite{Dietz1990} that an exact form written as an infinite sum is quite close to the approximate expression given by the Wigner surmise.

The relationship between the properties of the energy spectra of quantum systems that are strongly chaotic in the classical limit and the spectra of random matrices emerged in studies of billiards~\cite{Casati1980,Bohigas1984} and was supported with semiclassical analysis~\cite{GutzwillerBook}. In the other limit of completely integrable classical systems, it was understood~\cite{Casati1985} that the form of $P(s)$ can be approximately described by the Poisson distribution, $P(s) \sim \exp(-s)$.  This distribution was analyzed  within a semiclassical approach to quantum systems in~\cite{Berry1977}. The Berry-Tabor conjecture \cite{Berry1977} of the Poisson form of $P(s)$ is considered a generic property of integrable quantum systems with a classical limit, but despite intensive mathematical studies (see~\cite{Sinai1988, SinaiProceed, MarklofProceed} and references therein), it has not yet been rigorously proved.  Recent studies~\cite{Mailoud2021} of the Lieb-Liniger quantum model, known to be integrable and solvable with the Bethe ansatz \cite{Bethe1931,Korepin1993}, have shown that the Berry-Tabor conjecture fails due to the existence of underlying correlations between energy levels. Other works that have discussed the onset of level repulsion in integrable systems include Refs.~\cite{Benet2003,Relano2004,Scaramazza2016,Wang2022}.

As one moves from few- to many-body systems,  the analysis of the correspondence principle gets more challenging~\cite{Rammensee2018,Hummel2019,Engl2014,Waltner2017,Akila2017,Akila2018,Chan2018,Pozsgay2020,Schubert2021}. With regard to classical systems, this is primarily due to the structure of the multidimensional phase space, which becomes practically inaccessible for a detailed study. On the quantum side, the main problem is the dimension of the Hilbert space, which grows exponentially fast with the number of particles and sites. As a result, many questions about the QCC applied to many-body systems remain unanswered.

%%%%%%%%%%%%%%%%%%%%%%%%%%%%%%%%%%%%%%%%%%%%%%%%%%%%%%%%%%%%%%%%%%%%%%%%%%%

The purpose of this paper is to resolve the issue of the QCC for many-body systems, at least partially, by focusing on the strongly chaotic regime. Our study is based on the semianalytical approach that was used to establish the criteria for the onset of quantum chaos and statistical relaxation in isolated quantum systems of interacting Fermi and Bose particles, and which were confirmed by numerical experiments~\cite{Flambaum1997,Borgonovi2016,Borgonovi2017,Borgonovi2019}.

Our analysis is done for a one-dimensional spin model, that is relevant to experiments with ion traps, where the range of the interactions can be tuned~\cite{Richerme2014,Jurcevic2014,Smith2015,Brydges2019}. We show that quantum functions used in the analysis of quantum dynamics, quantum chaos, and localization have classical counterparts, and the QCC between them is excellent in the region of quantum chaos. We also find a direct relationship between the rate of quantum information spread and the Kolmogorov-Sinai entropy, which is a classical quantity obtained by summing the positive Lyapunov exponents.

%%%%%%%%%%%%%%%%%%%%%%%%%%%%%%%%%%%%%%%%%%%%%%%%%%%%%%%%%%%%%%%%%%%%%%%%%%%
The paper is organized as follows. In Sec.~\ref{Sec:Model}, we describe our one-dimensional classical system with $L$ interacting spins and determine the range of parameters chosen for our numerical simulations. As we explain, the analysis of the system's  behavior in the multidimensional phase space is simplified due to additional integrals of motion. The phase space of the system represents a set of $L$ three-dimensional (3D) spheres, each one corresponding to one spin. Therefore, the motion of each spin is limited by a 3D phase space and is described by a 3D equation with an external perturbation determined by the behavior of the remaining spins. As a result, the degree of instability of the motion of any spin is effectively determined by the single positive Lyapunov exponent associated only with each individual spin and not with all the spins of the system. This allows us to relate the maximal Lyapunov exponent with the Kolmogorov-Sinai entropy. In addition, we are able to obtain a rigorous and simple definition of ergodicity by examining the motion of each spin on its 3D sphere.

Section~\ref{Sec:Quantum}  describes the corresponding quantum system and gives an estimate of the critical interaction strength above which the behavior of the quantum system can be considered chaotic. The analysis is based on the structure of the Hamiltonian matrix represented in a properly chosen noninteracting basis. This allows us to perform a qualitative analysis of the properties of the system and to estimate the parameters for the emergence of chaos as determined by the energy spectrum of the system, in particular, the onset of the Wigner-Dyson distribution for the spacings between the nearest levels. Numerical data support the analytical estimates that we obtain.

In Sec.~\ref{Sec:LDOS} we show that it is possible to define classical analogues to  functions that are widely used in the analysis of quantum systems. One of these functions is known in nuclear physics as strength function and in solid state physics as local density of states (LDoS). The other refers to  the shape of the eigenfunctions (SoE) and is used to quantify the eigenstates as either localized or delocalized, and as regular or chaotic.  The LDoS and the SoE for classical systems  are obtained by numerically integrating the classical equations of motion~\cite{Borgonovi1998a,Izrailev2001}.  The possibility to talk about the classical LDoS and the classical SoE remains little known, despite some previous studies~\cite{Borgonovi1998,Luna2001,Luna2002,Benet2000,Benet2003b}. Our numerical data demonstrate an excellent correspondence between the classical and quantum functions above the quantum chaos border. 
The knowledge of these functions is extremely important for the description of the dynamical properties of the quantum system and its relaxation into a state that can be described statistically~\cite{Borgonovi2016}. 

In view of the QCC, several results can be obtained relating the behavior of quantum observables, such as survival probability, Loschmidt echo, different types of entropies, and OTOCs, with the maximal Lyapunov exponent, $\lambda_{\text{max}}$, in the region of strong chaos~\cite{Cucchietti2002,Gorin2006,Wijn2012,Rozenbaum2017,Chavez2019,Rautenberg2020,Wang2020,Wang2021}.  
However, the global dynamics of  many-body systems should be related with the Kolmogorov-Sinai entropy rather than the maximal Lyapunov exponent. This prompts us to search for a link between the quantum dynamics and the Kolmogorov-Sinai entropy, a direction taken also in~\cite{Miller1999,Asplund2016,Bianchi2018}.
It was recently shown that the width of the quantum LDoS determines the dynamical wave packet spreading in the non-interacting many-body Hilbert space~\cite{Borgonovi2019,Borgonovi2019b} and it is directly related with the timescale for equilibration. This motivates  Sec.~\ref{Sec:KS-LDOS}, where we compare the width of the LDoS with the Kolmogorov-Sinai entropy, finding a direct correspondence between the two.
Conclusions are given in Sec.~\ref{sec:conclusions}.

%%%%%%%%%%%%%%%%%%%%%%%%%%%%%%%%%%%%%%%%%%%%%%%
%%%%%%%%%%%%%%%%%%%%%%%%%%%%%%%%%%%%%%%%%%%%%%%

\section{Classical Model}
\label{Sec:Model}
We consider a one-dimensional classical model of $L$ interacting spins  described by the following Hamiltonian, 
\begin{equation}
\label{eq:ham1}
H = H_0 + V = \sum_{k=1}^L B_k S_k^z -
\sum_{k=1}^{L-1}   \sum_{j>k}^{L} J_{jk} 
S_j^x S_{k}^x ,
\end{equation} 
where
$$B_k \equiv \left(B_0+\delta B_k \right),$$
and
$$J_{jk} \equiv  \frac{J_0}{|j-k|^\nu}.$$ 
The angular momentum $I^2 = |\vec{S}_k|^2$, for $k=1,...,L$, is fixed. We choose  $I=1$, so that time has the dimension of inverse energy.
In Eq.~(\ref{eq:ham1}), $B_k$ are the frequencies of the noninteracting motion described by the Hamiltonian $H_0$. They are slightly detuned by random values of $\delta B_k$, with $|\delta B_k| \leq \delta W \ll B_0 $,  to avoid degeneracies in the corresponding quantum model. We consider a single set of random values of $\delta B_k$, so the model is perfectly deterministic, in the sense that no averages over different realizations of Hamiltonians are performed. The interacting part of the Hamiltonian depends on the couplings $J_{jk}$ between all pairs of spins and  decays algebraically with the distance between the spins with an exponent $\nu > 1$. Even though there is coupling between distant spins, technically speaking, since the model is one-dimensional and $\nu >1$, we are not considering long-range interactions. Most of our results are obtained for the generic value of $\nu=1.4$ and do not depend on this choice. 

The classical equations of motion are obtained from standard expressions written in terms of the Poisson brackets and the Levi-Civita symbol $ \epsilon^{\alpha \beta \gamma}$ as 
\begin{equation}
\left\{S_k^\alpha, S_j^\beta\right\}
= \delta_{kj} \epsilon^{\alpha \beta \gamma} S_k^\gamma , 
\end{equation}
 from which we have 
\begin{eqnarray}
\dot{S}_k^x &= &\left\{ S_k^x, H_0 \right\} = - B_k  S_k^y , \nonumber
\\
\dot{S}_k^y &=& \left\{ S_k^y, H \right\} =   B_k  S_k^x + S_k^z \sum_{j\ne k}  J_{jk} S_{j}^x , 
\label{eq:eqm}
\\
\dot{S}_k^z &=& \left\{ S_k^z, V\right\} =  - S_k^y \sum_{j\ne k}   J_{jk} S_{j}^x .\nonumber
\end{eqnarray} 

The numerical solution of the classical dynamics is obtained by integrating, via the 9/8 Runge-Kutta algorithm, the equations of motion in Eq.~(\ref{eq:eqm}). To make the comparison between quantum and classical dynamics as close as possible, the initial conditions for the classical system are chosen using random directions for the spins in the Bloch sphere under either the constraint of (i)  fixed noninteracting energy in a sufficiently small microcanonical energy range $E_0-\delta E_0 < H_0 < E_0 +\delta E_0 $,  or of (ii) fixed total energy $E-\delta E < H < E+\delta E$. For each result, we specify  whether condition (i) or (ii) is used.

When analyzing  the structure of the phase space of model (\ref{eq:ham1}), it is important to take into account that in addition to the total energy, there are  $L$ other integrals of motion, namely the squares of the angular momenta of each spin, which are fixed. In other words, the classical trajectory of each individual spin lies on a sphere of unit radius, which is separate from the other spins. The influence of the surrounding spins to  the stability of an individual spin can be regarded as an external perturbation. From this viewpoint, the  equation for the $k$-th  spin can be written as
\begin{equation}
\label{eq:po}
\ddot{S_k^z} + \Omega_k^2(t) S_k^z = F_k(t),
\end{equation}
with the nonlinear time-dependent frequency 
\begin{equation}
\label{om}
\Omega_k^2 (t) = \left[ \sum_{j\ne k} J_{jk} S_j^x(t) \right]^2 = J_0^2 \left[\sum_{j\ne k} 
\frac{ S_j^x(t)}{|j-k|^\nu} \right]^2 ,
\end{equation}
and the driving nonlinear force
\begin{eqnarray}
\label{drive}
F_k (t) = \sum\limits_{j\ne k} J_{jk} \left[B_j S_j^y(t) S_k^y(t)- B_k S_j^x(t) S_k^x(t)\right] \\
= J_0 \sum\limits_{j\ne k} \frac{ B_j S_j^y(t) S_k^y(t)- B_k S_j^x(t) S_k^x(t) } {|j-k|^\nu} .\nonumber
\end{eqnarray}
Equation~(\ref{eq:po}) indicates that the $z$-component of each single spin can be thought of as a parametric oscillator with a time-dependent frequency $\Omega_k(t)$ and under the force $ F_k(t)$ that depends on the sum of the product of all $x$ and $y$ spin components. 

The picture above implies that, in addition to the stability of the motion of the total system consisting of $L$ oscillators (spins), one can ask about the stability of any individual spin. 
Notice that the motion of a single spin is itself  nonlinear, but for weak interactions between spins, the nonlinear terms can be treated perturbatively. This representation of our model helps to understand the essential properties of the motion. Specifically, in the weak interaction limit,  the motion of the $S_z$ component of a chosen oscillator can be considered separately from the $S_x$ and $S_y$ components. 

A somewhat similar model of  parametric oscillators was analyzed by Chirikov in his seminal paper \cite{Chirikov1960a,Chirikov1960b}, in which he discovered that the overlaps of nonlinear resonances result in the phenomenon of chaotic motion. These resonances appear in the second order of perturbation theory, which complicates the analytical approach. The rigorous analysis of such  {\it nearly-linear} models of interacting particles, where the nonlinearity is due to the perturbative coupling with other degrees of freedom, is still an open problem (see \cite{Izrailev1980}).

In the following two subsections we study the  properties of our system with the purpose of identifying, if any, the region of maximal classical chaos. We also give a physical insight of what maximal chaos actually means. 

\subsection{Lyapunov Analysis}
\label{Sec:Lyapunov}

\begin{figure}[t]
    \centering
    \includegraphics[width=4.25cm]{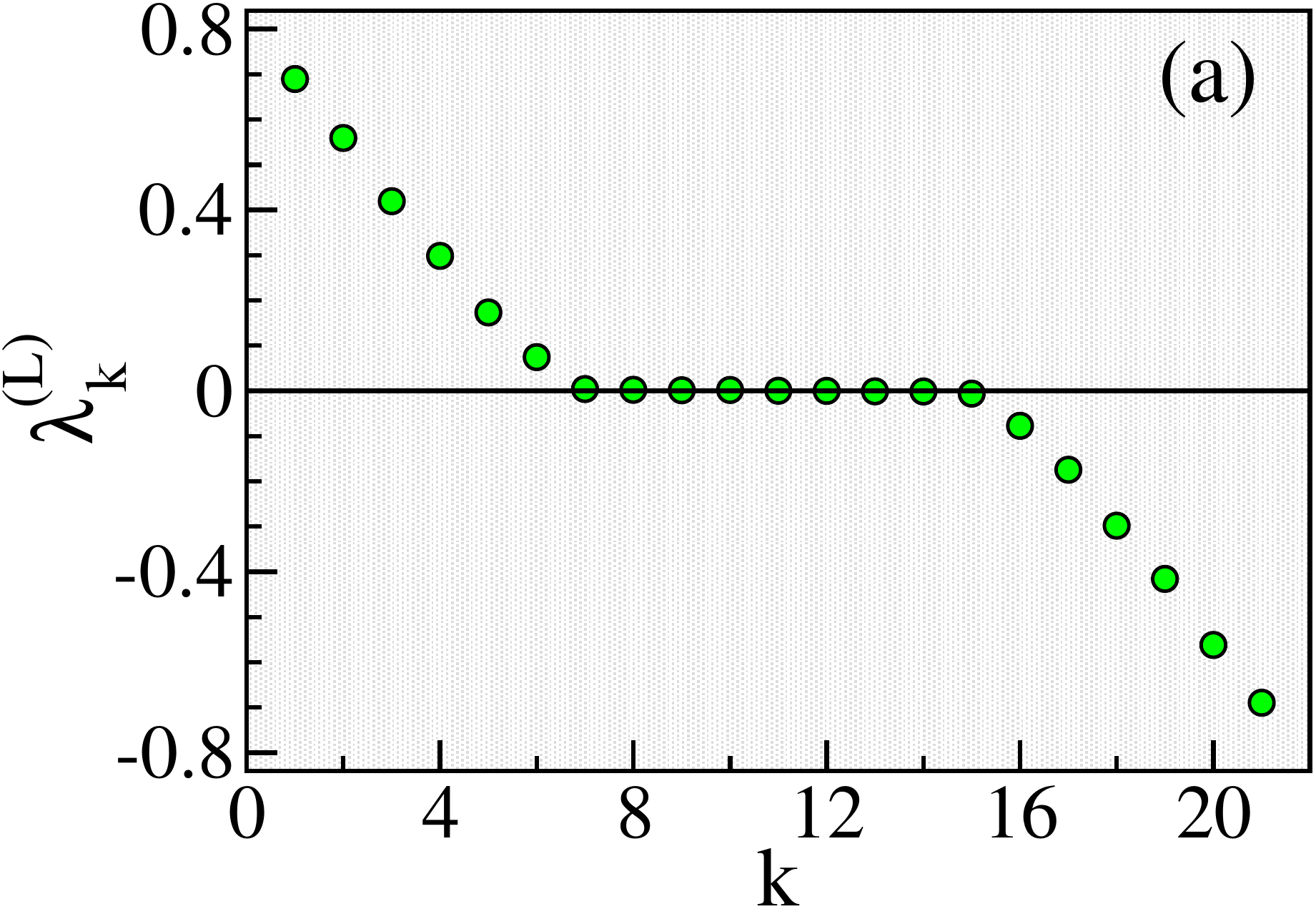}
    \includegraphics[width=4.25cm]{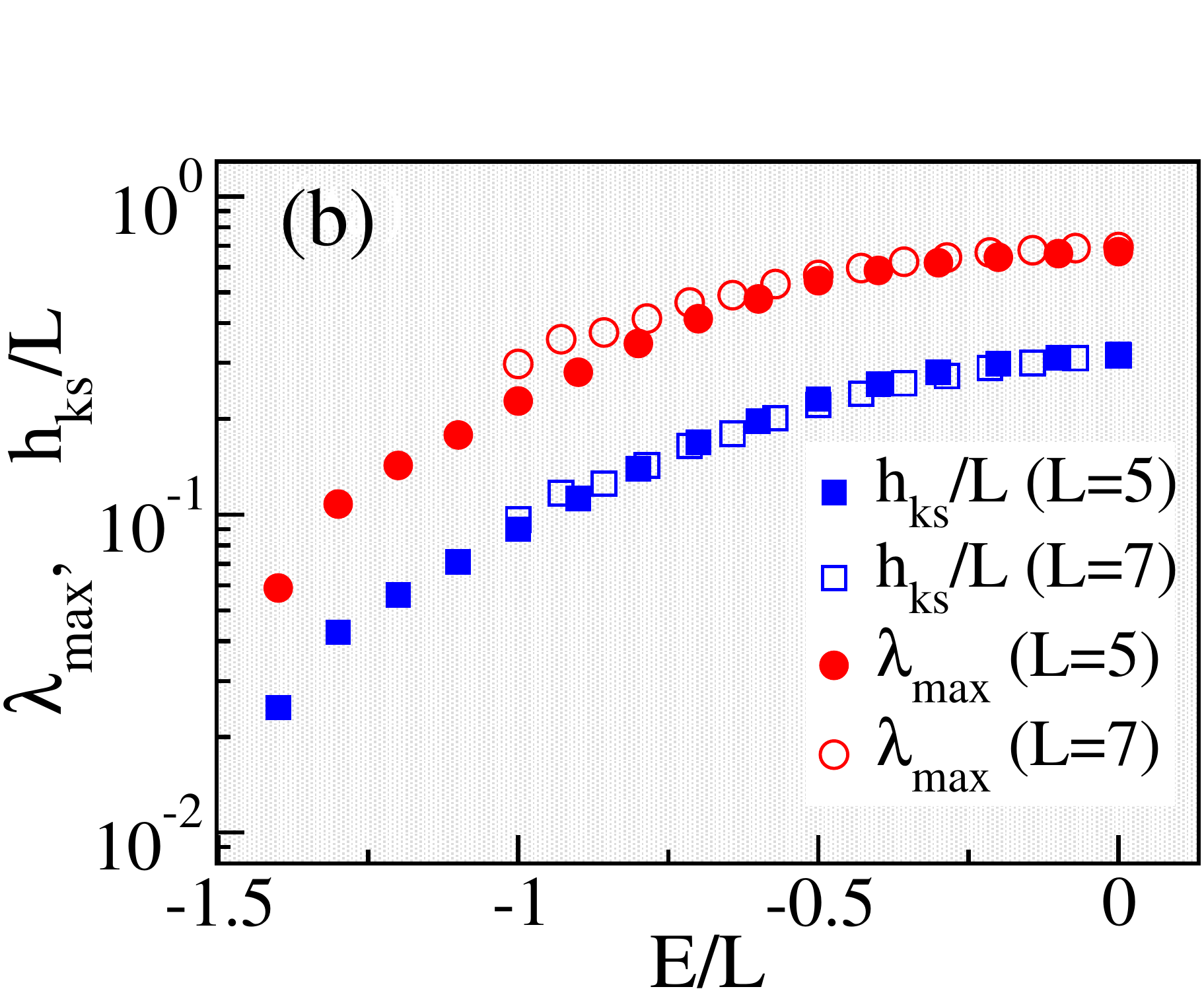}
    \\
    \includegraphics[width=4.25cm]{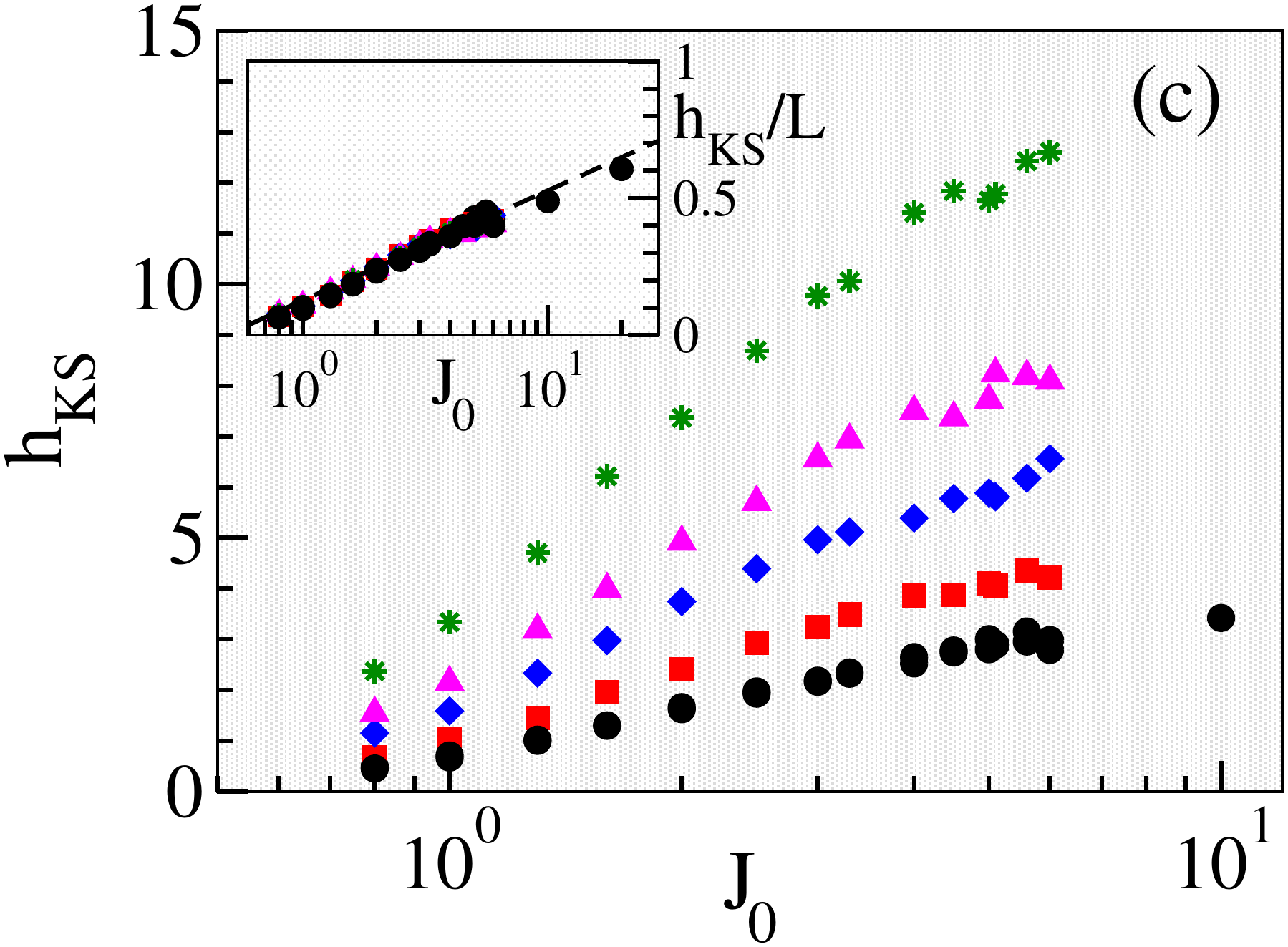}
    \includegraphics[width=4.25cm]{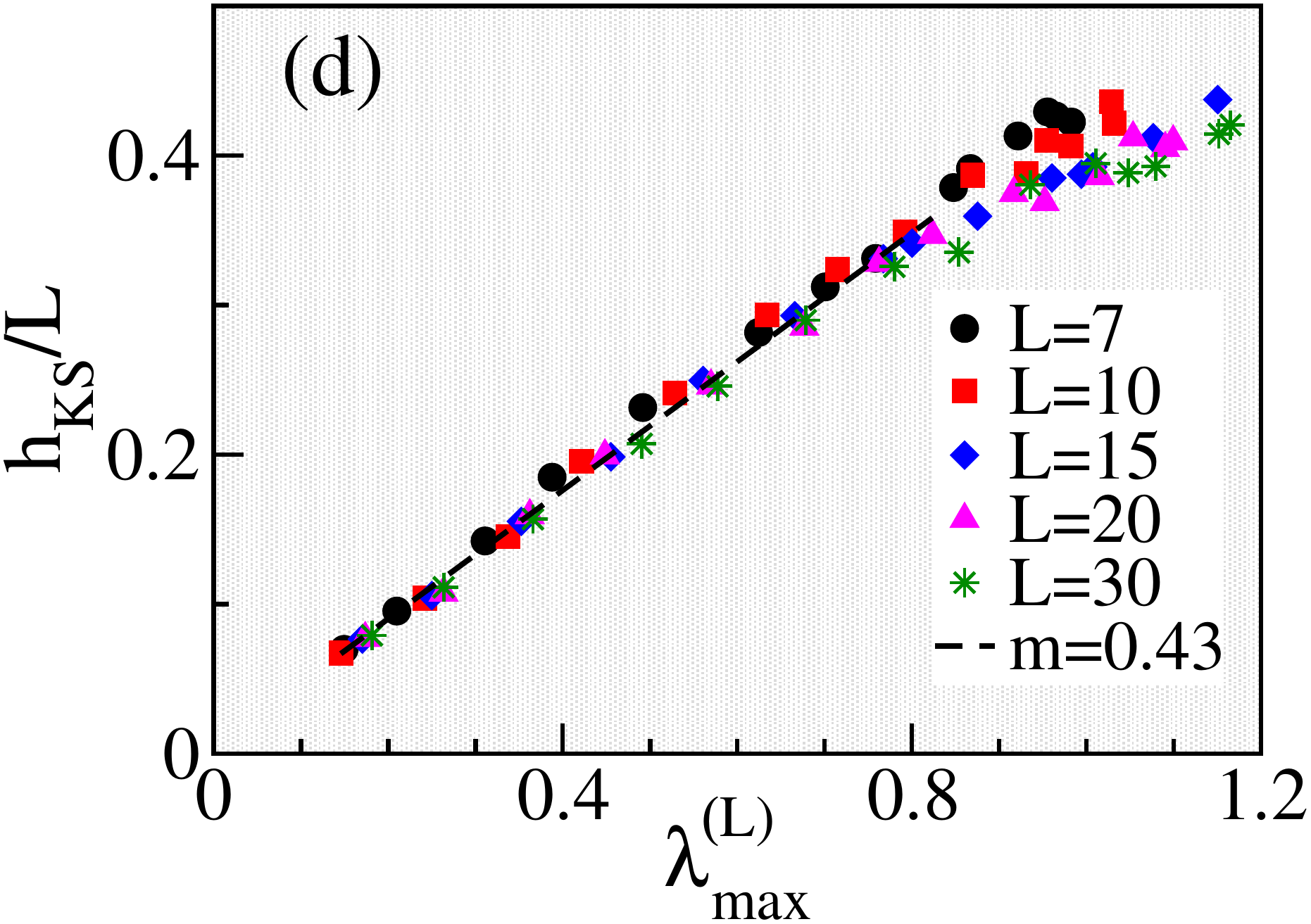}
    \caption{(a) Lyapunov spectrum for $L=7$ spins showing $L$ positive, $L$ negative, and $L$ null exponents. 
    (b) Average (over initial conditions) of the maximal Lyapunov exponent and average density of the Kolmogorov-Sinai entropy  as a function of the energy per spin, $E/L$, for two different system sizes $L$.
    (c) Average Kolmogorov-Sinai entropy as a function of $J_0$ (in semilog scale) for different values of $L$ [see legend in (d)]. Inset: rescaled Kolmogorov-Sinai entropy as a function of $J_0$ showing the collapse into a single curve. 
    (d) Average Kolmogorov-Sinai entropy density {\it vs } average maximal Lyapunov exponent for different values of $L$; same values of $J_0$ as in (c). Dashed line is the best linear fit ($m$ is the slope) for all data in the region $\lambda^{(L)}_{\text{max}} < 0.8$. Deviations from the linear dependence appear for  $\lambda^{(L)}_{\text{max}}>0.8$, since this region is characterized by large values of the interaction strength, $J_0 \gg B_0$.
    In all panels: $B_0=1$, $\delta W=0.2$, and $\nu=1.4$. In (a)-(b): $J_0=3$. In (b): Average over an ensemble of 300 initial random trajectories within a small energy window $\pm 0.01$ around the indicated energy values in the $x$-axis. In (c)-(d): Fixed energy $|E|<0.01$ and the average is done over $10^3$ initial values of $S^{x,y,z}$.  }
    \label{fig:lyaphks}
\end{figure}

To study the chaotic properties of the classical many-body system, we perform the standard Lyapunov analysis, which consists in finding the Lyapunov spectrum $\lambda_1, \lambda_2, \ldots \lambda_{3L}$ \cite{Iubini2021}. This is done by choosing random initial conditions at a fixed energy $E$. Since the length of each spin is a constant of motion,  $L$ exponents are equal to zero.  In addition, due to the time-reversal symmetry, the eigenvalues satisfy $\lambda_k=-\lambda_{3L-k}$, for $k=1,..L$, hence there are only $L$ positive Lyapunov exponents. The spectrum is illustrated in Fig.~\ref{fig:lyaphks}~(a) for $L=7$. Notice that the Lyapunov exponents corresponding to $k=7$ and $k=15$ are very close to zero, but are not zero.

%%%%%%% FIG 01 %%%%%%
%%%%%%%%%%%%%%%%%%%%% 

From the Lyapunov spectrum, we  obtain the Kolmogorov-Sinai entropy, $h_{\text{KS}}$, which is defined via the Pesin theorem~\cite{Pesin1977}  as the sum of all positive Lyapunov exponents,
 \begin{equation}
     \label{eq:hks}
     h_{\text{KS}} = \sum_{k=1}^L \lambda_k^+ .
 \end{equation}
Our main interest is in the maximal Lyapunov exponent, $\lambda_{\text{max}}$.  For two sufficiently close initial conditions on the constant energy surface, the distance between the two trajectories in the multidimensional phase space grows exponentially in time with a rate given by $\lambda_{\text{max}}$. Its inverse, $1/\lambda_{\text{max}}$,  defines the shortest time-scale related to the dynamical instability. 

Generically, the Lyapunov exponents and the Kolmogorov-Sinai entropy depend on the initial conditions, such as the energy or the particular regions of the energy surface, and on the chosen parameters $L$, $B_0$, $J_0$, and the range of the interaction $\nu$. To show this dependence, we fix $B_0=1$, which sets the energy scale, keep $\nu=1.4$, choose a sufficiently large interaction strength $J_0=3$,  and study in Fig.~\ref{fig:lyaphks}~(b) $\lambda_{\text{max}}$ and $h_{\text{KS}}$ as a function of the energy of the initial conditions. Here and hereafter, the notation for the maximal Lyapunov exponent, $\lambda_{\text{max}}$, and for the Kolmogorov-Sinai entropy, $h_{\text{KS}}$, indicate the average values obtained within an ensemble of random initial conditions with constant energy $E$. As seen in Fig.~\ref{fig:lyaphks}~(b), both the maximal Lyapunov exponent and the Kolmorogov-Sinai entropy are smooth increasing functions of the energy density $E/L$. We therefore choose $E=0$ as the region of maximal dynamical instability (chaos) for our further investigations. In the numerical simulations $E=0$ means $ -0.01 < E < 0.01$.

It is important to note that fixing the value of the energy does not fix the degree of chaos, since we still have the freedom to tune the interaction strength $J_0$.   This is actually a subtle point that deserves better clarifications. Contrary to the common intuition,  increasing the inter-spin interaction strength, $J_0$, while keeping the energy fixed at $E=0$, does not increase the contribution of the interacting part $V$ in comparison with the non-interacting part $H_0$. Fixing the initial conditions  to have  $E=0$ means that $|H_0| \simeq |V|$, so for any $J_0$, both $H_0$ and $V$ remain on the same order of magnitude. Physically, changing the strength of the perturbation $J_0$ while keeping the energy $E$ fixed  just means exploring different regions of the energy surface $E=0$. That is, increasing  $J_0$ corresponds to selecting  a set of initial conditions where, on average, the modulus of the $z$-magnetization, $|(1/L) \sum_k S_k^z|$, also increases. This is a generic feature found whenever one fixes the total energy rather than changing the ratio between the interacting
part ($V$) and the non-interacting part ($H_0$) of the Hamiltonian.

Figure~\ref{fig:lyaphks}~(c) shows that the  Kolmogorov-Sinai entropy grows logarithmically with respect to the interaction strength $J_0$, even for small $J_0 < 1$. To better understand the dependence of $h_{\text{KS}}$ on the system size $L$,  we plot in the inset the density of the Kolmogorov-Sinai entropy, $h_{\text{KS}}/L$, as a function of $J_0$.  As one can see, the data obtained for different values of $L$ collapse into a single curve, showing that the rescaling  with respect to the system size is excellent. These results should be compared with those contained in~\cite{Iubini2021}, where an analogous rescaling was found for the discrete nonlinear Schr\"odinger chain.

%%%%%%%%%%%%%%%%%FIG 2  
 \begin{figure}[t]
    \centering
    \includegraphics[width=8cm]{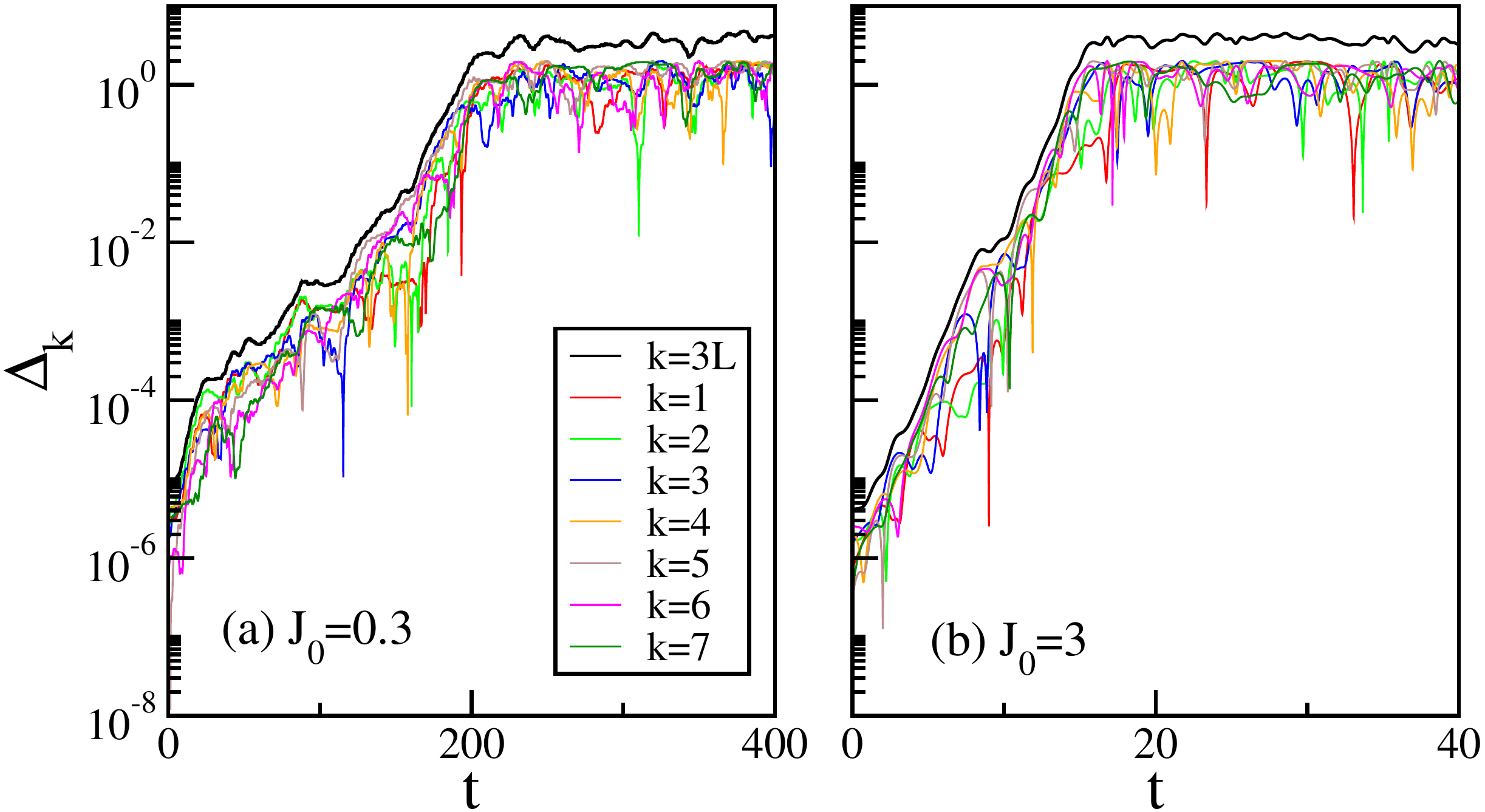}
    \caption{Distance between two close trajectories in time. Black thick line is the Euclidean distance in the $3L$ dimensional space, while colored lines stand for the distance in the 3D space for different spin numbers $k=1,..,L$, where $L=7$.
    The parameters are $\alpha=1.4$, $B_0=1$, $\delta W=0.2$, initial energy $E=0\pm 0.01$, and interaction strength $J_0=0.3$ below the chaos border in (a) and  $J_0=3$ above the chaos border in (b). Only one trajectory is considered.
    }
    \label{fig:lydy}
\end{figure}
%%%%%%%%%%%%%%%%%%%%%%%%%%%%

For values of $J_0$ that are not too large with respect to $B_0$, the (average) maximal Lyapunov exponent shows a dependence on the interaction strength similar to that of the density of Kolmogorov-Sinai entropy. To facilitate the comparison, we plot one quantity as a function of the other in Fig.~\ref{fig:lyaphks}~(d). There is a linear relation between them,  
$h_{\text{KS}}/L = m  \lambda_{\text{max}}$, where $m \simeq 0.43 $ is the slope obtained by the best linear fitting for not too large values of $\lambda_{\text{max}}$. This  relationship is in  agreement with the approximate linear dependence of the positive part of the Lyapunov spectrum as a function of $k$ seen in Fig.~\ref{fig:lyaphks}~(a).

Concerning the physical meaning of these results, let us first point out that
the dependence of both $h_\text{KS}$ and $\lambda_\text{max}$ on $\ln J_0$, as evident from Fig.~\ref{fig:lyaphks}~(c) and Fig.~\ref{fig:lyaphks}~(d), raises the question of the contribution of each spin to the growth in time of the phase space volume. Indeed, since a coarse-grained volume ${\cal V}$ in the phase space is expected to grow exponentially in time with an exponent  given by $h_\text{KS}$ \cite{Zaslavsky1981,ZaslavskyBook}, we should have
\begin{equation}
\label{eq:Zy}
 {\cal V}  (t) = {\cal V} (0) \ e^{h_\text{KS} t } \propto [J_0^{L}]^t.   
\end{equation}
The equation above indicates that even in the presence of a strong inter-spin interaction, each spin  contributes individually to the growth of this volume. This behavior must be related with the presence of the $L$ constants of motion, being therefore generic for spin systems. It would be interesting to test this result  in other classical spin chains.

Another important point that we raise, motivated by the equations of motion [Eq.~(\ref{eq:po})] for the single spins,  is the relationship between the maximal Lyapunov exponent of the multi-dimensional problem and the dynamical instability of the trajectory of each individual spin confined to its 3D-sphere.  To address this question, we compare the dynamics of the Euclidean distance between two initially close trajectories for each spin $k$, 
 $\Delta_k (t) = || \vec{S}_k (t)- \vec{S}^\prime_k (t)|| $, with the Euclidean  distance for the many-dimensional problem, $\Delta_{3L} = [\sum_{k=1}^L || \vec{S}_k (t)- \vec{S}^\prime_k(t)||^2]^{1/2}$. The results are shown in Fig.~\ref{fig:lydy} for two values of the interaction strength $J_0$. Despite fluctuations, one sees that the exponential growth of $\Delta_k (t)$ for each spin $k$ coincides with the maximal Lyapunov exponent of the many-body spectrum. This means that the timescale for the dynamical instability on each individual Bloch sphere is also described by the maximal Lyapunov exponent of the multidimensional problem, a result that is far from trivial.

 \subsection{Ergodicity of the Classical Motion}
\label{Sec:erg}

The finding above that the motion of each spin is characterized by the maximal positive Lyapunov exponent means that the motion is {\it locally } strongly chaotic. However this does not imply that global properties, such as the relaxation of the whole system of $L$ spins, follow the predictions of statistical mechanics. For this reason, we now investigate whether our system exhibits ergodicity. Apart from simple specific systems, such as billiards and two-dimensional maps, a mathematical proof of ergodicity for interacting many-body  systems is still missing. The usual numerical approach is to search for consequences of ergodicity, such as the decay of correlations functions, which is still not enough to claim ergodicity.

   %%%%%%%%%%%%%%%%%%%%%FIG 3 
\begin{figure}[t]
    \centering
    \includegraphics[width=0.45\textwidth]{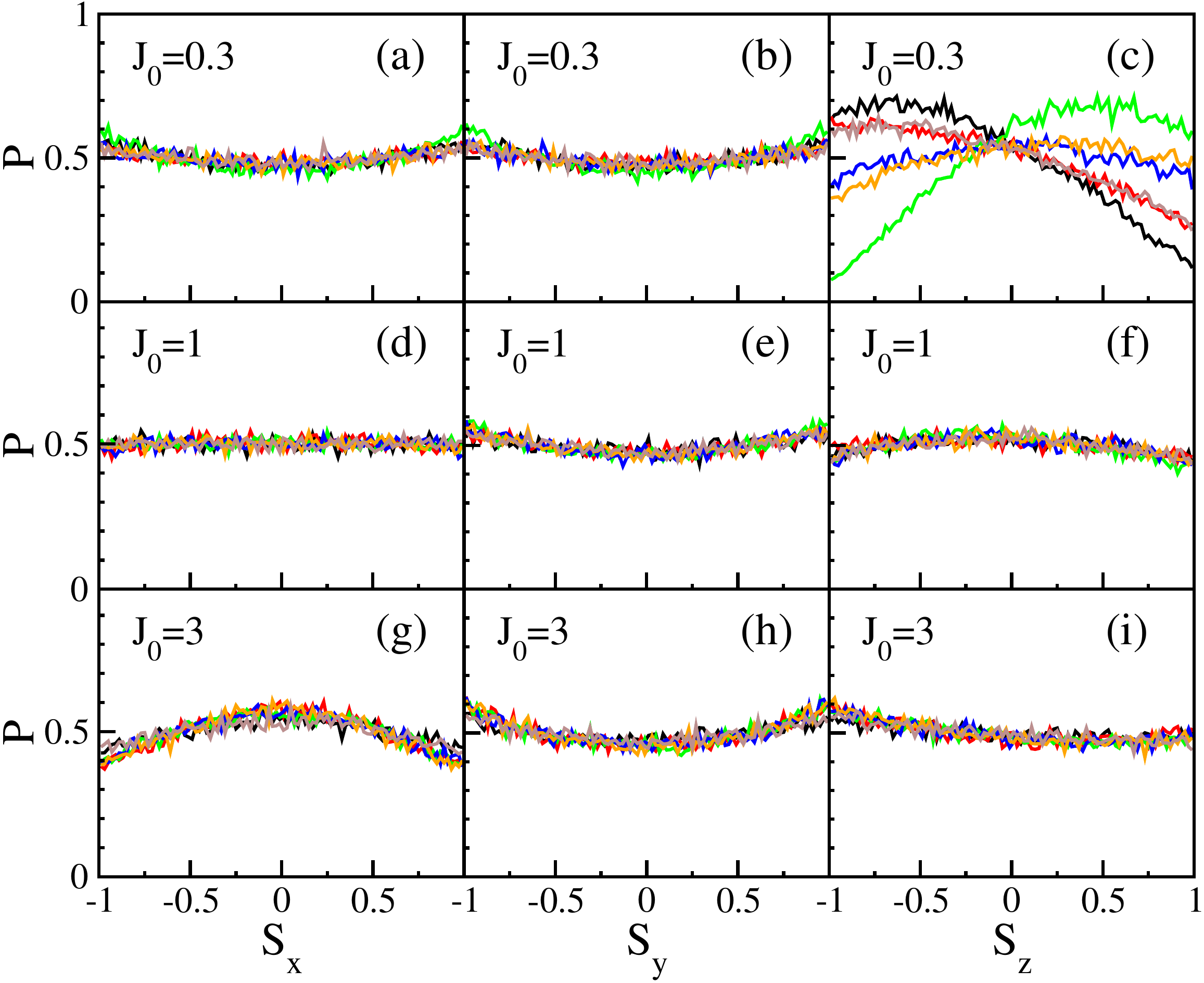}
    \caption{ Probability distribution functions $P(S_x)$ [(a), (d), (g)]; $P(S_y)$ [(b), (e), (h)], and $P(S_z)$ [(c), (f), (i)] for  $J_0=0.3$ [(a)-(c)], $J_0=1$ [(d)-(f)], and $J_0=3$ [(g)-(i)].
    Parameters:  $L=6$, $B_0=1$, $\nu=1.4$, $\delta W=0.2$.
    Initial conditions: random spins with energy $E=0$. Integration time $T=10^5$ Only one trajectory has been considered.
}
    \label{fig:e6}
\end{figure}
%%%%%%%%%%%%%%%%%%%%%%FIG 4 
\begin{figure}[t]
    \centering
    \includegraphics[width=0.45\textwidth]{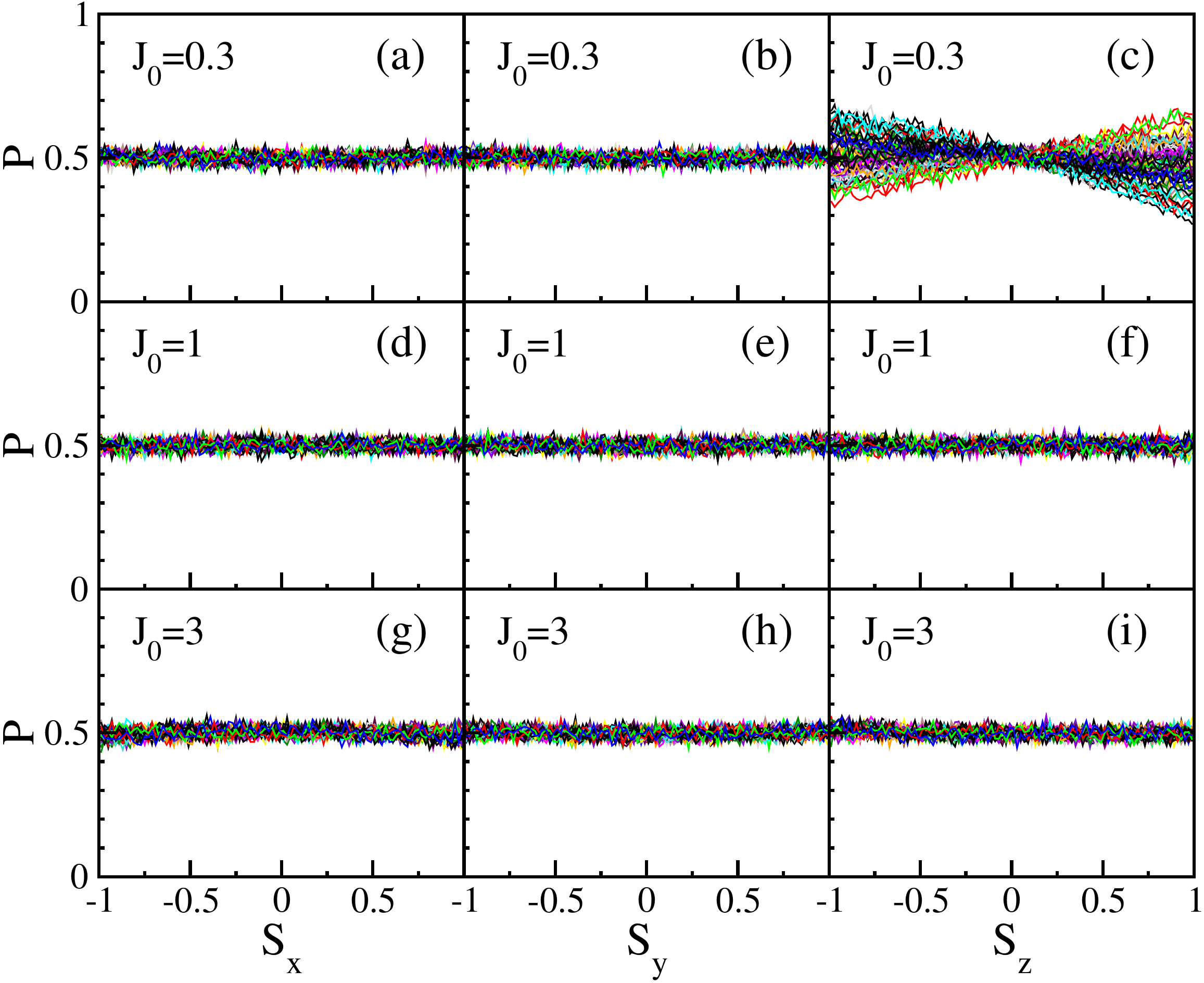}
    \caption{Probability distribution functions $P(S_x)$ [(a), (d), (g)]; $P(S_y)$ [(b), (e), (h)], and $P(S_z)$ [(c), (f), (i)] for  $J_0=0.3$ [(a)-(c)], $J_0=1$ [(d)-(f)], and $J_0=3$ [(g)-(i)].
    Parameters:  $L=60$, $B_0=1$, $\nu=1.4$, $\delta W=0.2$.
    Initial conditions: random spins with energy $E=0$. Integration time $T=10^5$.
 }
    \label{fig:e60}
\end{figure}
%%%%%%%%%%%%%%%%%%%%%%%%%%%%%
  
Here, we introduce a very simple and rigorous way to study the onset of ergodicity. Since the motion of each spin is confined to its 3D-sphere, a rigorous definition of ergodicity means that the distribution of the single spin Cartesian components follows the expression from the theory of $N$-dimensional full random matrices~\cite{Izrailev1990},
  \begin{equation}
      \label{eq:f90}
      P(S) = \frac{\Gamma(N/2)}{\sqrt{\pi}\Gamma((N-1)/2)}(1-S^2)^{(N-3)/2}
  \end{equation}
  where $\Gamma(N)$ is the Gamma function.
  In the case of $N \to \infty $, this expression leads to the Gaussian distribution of the eigenvectors components. Since in our case we have $N=3$,  the ergodic distribution implies the constant result $P(S)=1/2$. This method significantly simplifies the study of ergodicity for spin models.
  
To test ergodicity in our system, we consider a single initial condition $u(0) = \{ \vec{S}_1(0),..,\vec{S}_L(0) \} $ at a fixed energy $E=0$ and compute the trajectory $u(t)$ under the full Hamiltonian $H$ for a very long time $T$. We then build the distribution obtained by sampling  each component $S_k^\alpha (t_j)$, where $\alpha= x,y,z$ and $k=1,..,L$, at different times $t_j = j/T$, $j=1,..,T$. The results for the  distributions $P(S_x)$, $P(S_y)$, and $P(S_z) $ for various interaction strengths $J_0$ are shown in Fig.~\ref{fig:e6} and Fig.~\ref{fig:e60} for $L=6$ and $L=60$ respectively. Each color represents a different spin number $ k=1,...,L$.

In Figs.~\ref{fig:e6} (a)-(c), where chaos is weak, the distribution of $S_z$ is clearly different from that of the components $S_x$ and $S_y$. This can be understood from the equations of motion, Eq.~(\ref{eq:po}). In the absence of interaction, the motion of each spin is simply a rotation about the $z$-axis. In the presence of weak interaction strength $J_0$, these oscillators get coupled, resulting in a motion that covers ergodically a  portion of the surface of the 3D sphere. For even larger $J_0$, the motion eventually covers the whole surface.
Looking at Figs.~\ref{fig:e6}, as $J_0$ increases to $J_0=1$ [Figs.~\ref{fig:e6}~(d)-(f)] and $J_0=3$ [Figs.~\ref{fig:e6}~(g)-(i)], $P(S_z)$ becomes more similar to $P(S_x)$ and  $P(S_y)$, but the distributions do not get flat, so the motion cannot be considered truly ergodic for this system size ($L=6$), even when the interaction strength is large (by significantly increasing $J_0$, the distributions become even more curved).

By increasing the system size $L$,  our data in Fig.~\ref{fig:e60} show a clear onset of ergodicity provided the interaction strength is strong enough, $J_0 \gtrsim 1$. Therefore,  ergodicity for the motion of any single spin requires a large number of spins. A possible reason for the lack of ergodicity for $L=6$ is the presence of  stability islands due to non-linearity. This has been observed for $L=2$ in Ref.~\cite{Borgonovi1998} and references therein.
We reiterate that to have strong statistical properties one should have {\it both } strong chaos, signalled by a positive maximal  Lyapunov exponent, and ergodicity.

\section{Quantum Model}
\label{Sec:Quantum} 
 
Upon quantization we have, for each spin, 
$$
|\vec{S}^2| = \hbar^2 S(S+1) = I^2 ,
$$
where $S$ is the  quantum spin number and the $z$-component of the spin has the values $\hbar s$ with $s=-S, -S+1...,S-1, S$. For simplicity, we only consider integer spin numbers $S$. 
 
The classical limit is obtained by taking  the spacing between two levels of one component of the angular momentum  to zero. Since $I$ is a classical quantity with a fixed value, this means that we need to increase the number of levels. In other words, we introduce an effective Planck constant,
\begin{equation}
  \hbar = \frac{I}{\sqrt{S(S+1)}},
  \label{eq:heff}
\end{equation}
and the classical limit  is recovered for $S\to \infty$. 

We build the many-body Hilbert space of the total quantum Hamiltonian, 
\begin{equation}
 \hat{H} = \hat{H}_0 + \hat{V},
 \label{eq:totham}
\end{equation}
using the $z$-representation, where the basis $\ket{n}$ corresponds to the many-body eigenstates of the non-interacting Hamiltonian $\hat{H}_0$ and have non-interacting energies $E_n^{(0)}$,
\begin{equation}
    \label{eq:ham0}
    \hat{H}_0\ket{n}=  \sum_{k=1}^L \left(B_0+\delta B_k \right) \hat{S}_k^z \ket{n} \equiv  E_n^{(0)}\ket{n},
\end{equation}
with
$$ \ket{n} = |s_1,...,s_k,...,s_L\rangle, $$ 
and
$$
\hat{S}_k^z \ket{n} = \hat{S}_k^z |s_1,...,s_k,...,s_L\rangle = \hbar s_k |s_1,...,s_k,...,s_L\rangle.
$$ 
The interacting part of the classical Hamiltonian (\ref{eq:ham1}) is written in terms of operators as
\begin{equation}
    \hat{V} =  \sum_{j>k}^{L} \sum_{k=1}^{L-1} \frac{J_0}{|j-k|^\nu} \hat{S}_j^x \hat{S}_k^x. 
    \label{Eq:Vqu}
\end{equation}

The interparticle interaction is computed by taking into account that $\hat{S}_j^x= (\hat{S}_j^+ + \hat{S}_j^-)/2$ and 
\begin{equation}
    \begin{array}{lll}
        &\hat{S}_k^{\pm} |s_1,...,s_k,...,s_L\rangle  = \\
        &\\
         & \hbar \sqrt{S(S+1)- s_k(s_k\pm 1)} |s_1,...,s_k\pm 1,...,s_L\rangle .
    \end{array}
    \label{eq:spm}
\end{equation}
The two-body interaction in Eq.~(\ref{Eq:Vqu}) can be written as $\hat{V}=\hat{V}_{in} + \hat{V}_{out}$, where
\begin{equation}
\hat{V}_{in} =  \sum_{j\ne k} \frac{J_0}{4|j-k|^\nu} \left( \hat{S}_j^+ \hat{S}_k^- + \hat{S}_j^- \hat{S}_k^+ \right)
\label{Eq:Vin}
\end{equation}
and 
\begin{equation}
\hat{V}_{out} =  \sum_{j\ne k} \frac{J_0}{4|j-k|^\nu} \left( \hat{S}_j^+ \hat{S}_k^+ + \hat{S}_j^- \hat{S}_k^- \right).
\label{Eq:Vout}
\end{equation}
The term $\hat{V}_{in}$ couples spin configurations that have the same total magnetization along the $z$-axis, $\hat{M}_z= \sum_{k=1}^L \hat{S}_k^z$, so it does not change the value of the quantum number $\mu_z=-LS,-LS+1,...,LS-1,LS$.
The term $\hat{V}_{out}$ couples basis vectors that differ by two excitations, so it changes the total magnetization by a factor
of two, $\mu_z \rightarrow \mu_z\pm 2$. The eigenvalues can then be divided as belonging to an even (in units of $\hbar$)
or odd value of the total $z$-magnetization. The model is not integrable for any choice of the parameters $J_0 \ne 0 $, $B_0$, $\nu, \ne 0$.

\subsection{Hamiltonian Matrix Structure and Quantum Chaos Border}

Information about the structure of the Hamiltonian matrix is fundamental for a complete description of the system in terms of quantum chaos~\cite{Borgonovi2016}. From the structure of the matrix, one can estimate the quantum chaos border, that is, the interaction strength necessary  to produce chaotic eigenstates and thus relaxation and eventually  thermalization~\cite{Borgonovi2016,Berman2001}. The full analysis of the Hamiltonian matrix, eigenvalues, and eigenstates provides a more complete picture than relying only on the statistics of the spacings between neighboring energy levels. 

\begin{figure}
    \centering
     \includegraphics[width=0.53\textwidth]{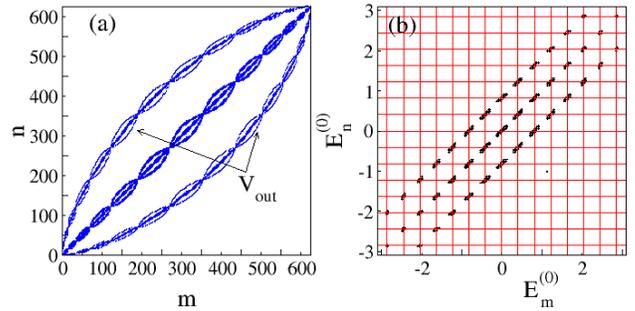}
     \vspace{-2.8cm}
    \caption{Structure of the full  Hamiltonian in the site representation (a) and in the energy representation (b).
    The block diagonal structure is evident. In (b), the red vertical and horizontal lines indicate multiples of $\hbar B_0$. For both panels: $B_0=1$, $\delta W=0.04$, $J_0=3$, $\nu=1.4$, $L=4$, $S=2$.
}
    \label{fig:mat}
\end{figure}

We order the many-body noninteracting states from low to high energies and analyze the structure of the Hamiltonian matrix in two representations. The site representation is used in Fig.~\ref{fig:mat}~(a), where the label $n$ in the axis indicates a many-body noninteracting state $|n\rangle$ and each colored dot marks a nonzero element $\langle n | \hat{H} | m\rangle \ne 0$. In Fig.~\ref{fig:mat}~(b), each  dot indicates again an element $\langle n | \hat{H} | m\rangle \ne 0$, but the labels in the $x$- and $y$-axes are now the noninteracting energies $E_n^{(0)}$. This noninteracting energy representation is more physical and it allows for generalizations to other many-body models.

To provide a detailed analysis of the Hamiltonian matrix, let us discuss first its noninteracting part $\hat{H}_0$, which commutes with $\hat{M}_z$.  This means that in the many-body $S_z$-representation the $H_0$ matrix has a block-diagonal structure with $2SL+1$ blocks, each block being associated with a quantum number $\mu_z$. The block structure is indeed observed for the diagonal elements in Fig.~\ref{fig:mat}~(a) and Fig.~\ref{fig:mat}~(b).

All levels belonging to a single block would be degenerate if $ \delta B_k = 0$. Instead, we have random $|\delta B_k| \leq \delta W \ll B_0 $, so each block of noninteracting levels forms an energy band whose maximal width is estimated to be $\approx 2L \delta W$. This can be confirmed by considering the central block with $\mu_z=0$ (which is the largest block)  and a generic state $\ket{n_0} = \ket{s_1,...,s_L}$ belonging to it. The noninteracting energy of this state is given by
$$
\bra{n_0} \hat{H}_0 \ket{n_0} = \hbar \sum_k s_i \delta B_k ,
$$
which is maximized for $s_k=S$ and $\delta B_k = \delta W$, so that $E^{(0)}_{max} = \hbar SL \delta W$,  
$E^{(0)}_{min} = -\hbar SL \delta W$, and the energy size of the central block can be estimated as
\begin{equation}
    \Delta_{0} =E^{(0)}_{max}- E^{(0)}_{min} \simeq 2\hbar SL\delta W \simeq 2L \delta W.
    \label{Eq:DeltaE}
\end{equation}
Since the energy distance between the diagonal blocks is proportional to  $\hbar B_0$  [see the horizontal and vertical red lines in Fig.~\ref{fig:mat}~(b)],   when  $B_0  > 2L S \delta W$,  the diagonal elements are arranged in $2LS+1 $ disconnected segments, as indeed seen in Figs.~\ref{fig:mat}~(a)-(b), while for $B_0 \lesssim  2LS \delta W$, all elements along the diagonal become more or less homogeneously distributed.

 %%%%%%% FIGS 05 %%%%%%

\begin{figure}[h]
    \centering
    \includegraphics[width=0.3\textwidth]{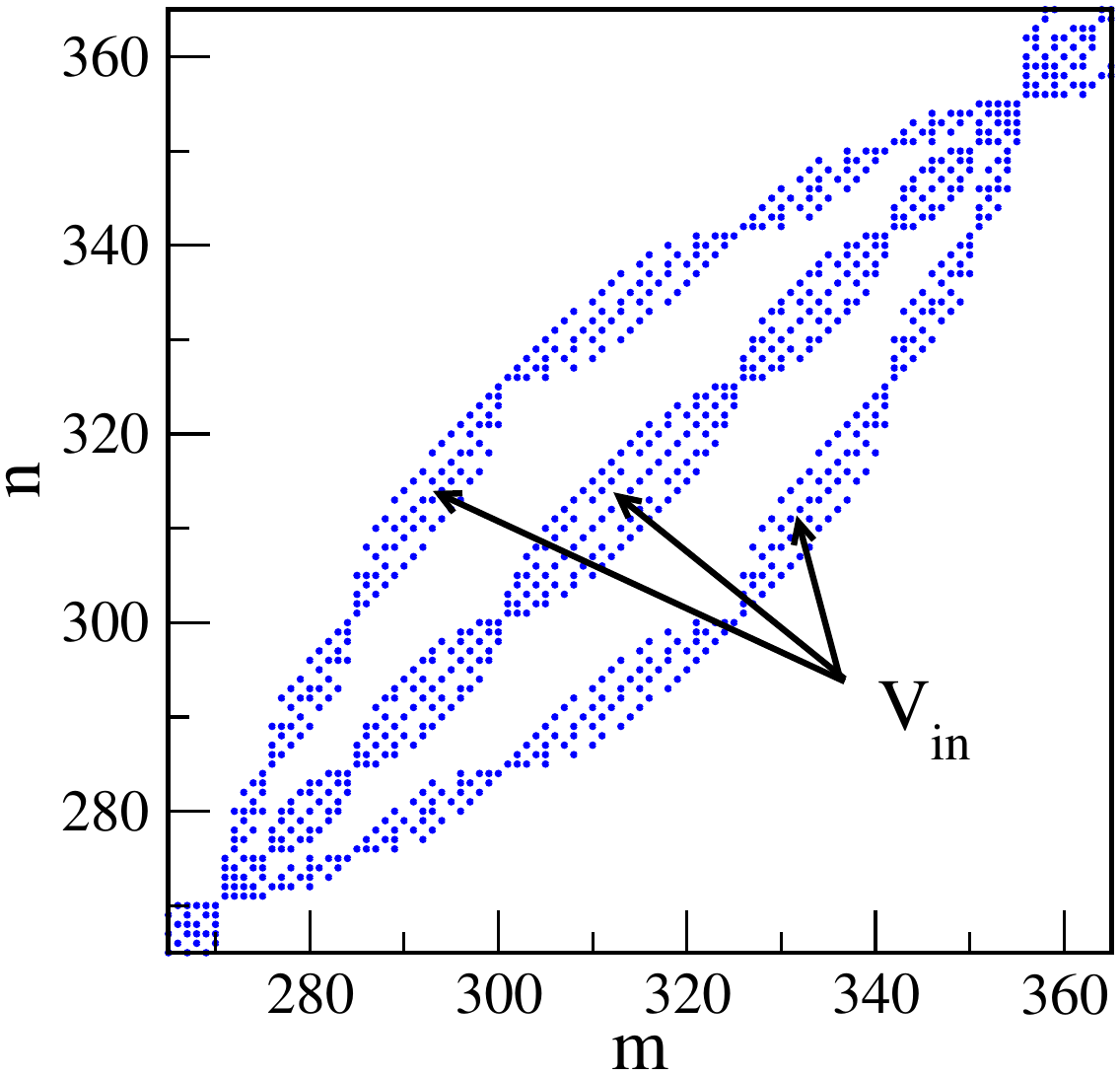}
    \caption{ Structure of the central block ($\mu_z=0$) of the  Hamiltonian $\hat{H}_0 + \hat{V}_{in}$  in the site representation.
    $B_0=1$, $\delta W=0.04$, $J_0=3$, $\nu=1.4$, $L=4$, $S=2$.  
    }
    \label{fig:mat1}
\end{figure}

We now move to the interacting part of the Hamiltonian, starting with the term $\hat{V}_{in}$ in Eq.~(\ref{Eq:Vin}) that only couples states inside the same block of fixed $\mu_z$-value, since it commutes with $\hat{M}_z$. Let us then concentrate on a single block with fixed $\mu_z$.
In Fig.~\ref{fig:mat1}~(a),  we show the central block ($\mu_z=0 $) and the arrows indicate examples of nonzero elements caused by the interactions within that block, $\langle n |\hat{V}_{in} | m\rangle \ne 0$.

%%%%%%%%%%%%%%%%%%%%%%%%%%%%%%%%%%%%
%%%%%%%%%%%%%%%%%%%%%%%%%%%%%

The two-body interaction $\hat{V}_{in}$, which contains the sum of the operators $ \hat{S}_j^+ \hat{S}_k^-$, exchanges the state of two spins that at most have a non-interacting  energy  difference $\delta \epsilon = \hbar |B_k -B_j| \simeq  2\hbar \delta W \simeq  2\delta W/S$.
 
The term $\hat{V}_{out}$ in Eq.~(\ref{Eq:Vout}) couples the elements of one block with those of a next-nearest neighboring block, and as such, it allows us to compute the entire width of the Hamiltonian matrix. The arrows in Fig.~\ref{fig:mat}~(a) indicate examples of nonzero elements caused by the interactions between the central block and outer blocks, $\langle n |\hat{V}_{out} | m\rangle \ne 0$. To estimate the energy bandwidth of the total Hamiltonian matrix, we sum the energy separation between the two outer bands connected with the $\mu_z=0$ central band, that is $\approx 4\hbar B_0$  
with the width of the outer band $2L \delta W$: $\Delta E \simeq 4B_0/S + 2L\delta W$.
 
In hands of the bandwidth of the Hamiltonian matrix, one now needs the number of directly coupled states, so that dividing that bandwidth by this number, one obtains the many-body energy spacing.   
To induce quantum chaos, the interaction strength has to be larger than the energy spacing, {\em i.e.} the chaotic regime emerges when the interaction is strong enough to mix all neighboring many-body levels. 

\begin{figure}[h]
    \centering
    \includegraphics[width=8 cm]{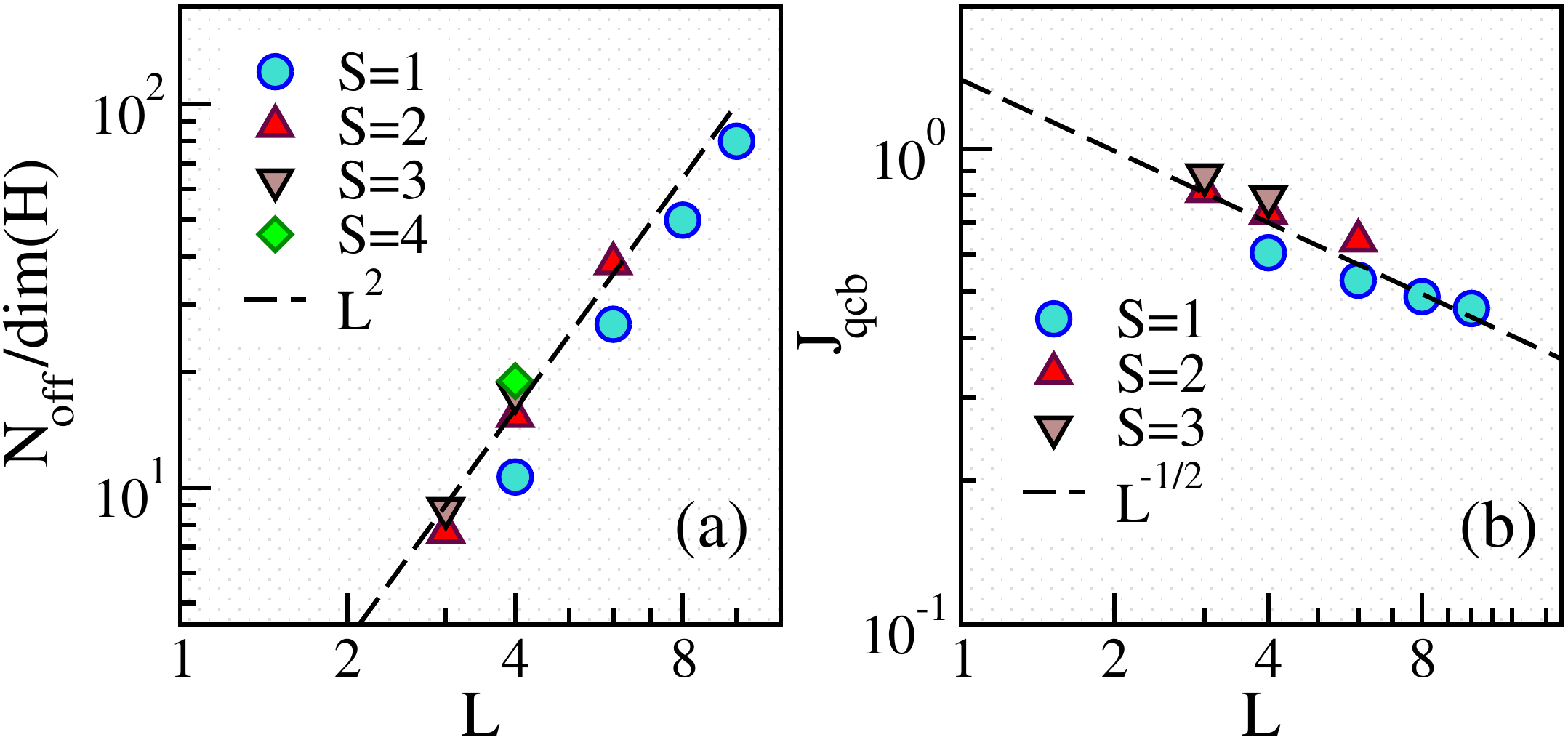}
    \caption{(a) Number of off-diagonal matrix elements $N_{\text{off}}$ rescaled by the dimension of the many-body Hilbert space $\text{dim}(H) = (2S+1)^L$ as a function of the system size $L$ for different values of $S$. The dashed line is $\propto L^2$. 
    (b) Interaction strength for the quantum chaos border $J_{qcb}$ as a function of $L$ for different values of $S$. The dashed  line is  $\propto 1/\sqrt{L}$. The parameters are $B_0=1$, $\delta W=0.2$, $J_0=1$, $\nu=1.4$. 
}
    \label{fig:voff}
\end{figure}

The average number of directly coupled states corresponds to the average number of nonzero off-diagonal elements in each row of the Hamiltonian matrix. This number should be proportional to the number of local operators in the interacting part of the Hamiltonian, $\hat{V}$, that is $\propto L^2$. Numerically, we compute
 the total number of off-diagonal matrix elements $N_{\text{off}}$ of the full Hamiltonian matrix and divide it by the matrix dimension, $\text{dim}(H) = (2S+1)^L$. This is done in Fig.~\ref{fig:voff}~(a) for different values of $L$ and $S$, confirming that indeed $N_{\text{off}}/\text{dim}(H) \propto L^2$. 

%%%%%%%%%%%%%%%%%%%%%%%%%%%%%
%%%%%%%%%%%%%%%%%%%%%%%%%%%%%

The quantum chaotic regime appears when the perturbation (strength of off-diagonal elements) is able to mix the neighboring many-body levels, that is,  $J_0 |V_{j,k}|  \gtrsim \Delta E/L^2 $, where $|V_{j,k}|$ represents the additional dependence of the values of the off-diagonal elements besides $J_0$, which includes $1/|j-k|^{\nu}$ and the terms from Eq.~(\ref{eq:spm}). Equivalently, one can write that quantum chaos emerges when
\begin{equation}
 J_0   \gtrsim  J_{qcb} = \frac{\Delta E}{L^2 |V_{j,k}|} ,
\label{eq:chbo}
\end{equation}
where $J_{qcb}$ is the interaction strength right at the quantum chaos border, a value that depends on parameters such as $\nu$, $L$, $S$.

In Fig.~\ref{fig:voff}~(b), we  show $J_{qcb}$ as  a function of the system size for different values of $S$. For the parameters used in our numerical investigations,  $S=1,2,3$, $B_0=1$, $\delta W=0.2$, $\nu =1.4$, $L=5,6,7$, an operative estimate for the quantum chaos border indicates $J_0 \gtrsim 0.5$. This is the quantum chaos threshold used in this work. However, as our Fig.~\ref{fig:voff}~(b) indicates and as it has been seen in  previous numerical studies~\cite{Santos2010PRE,Torres2014PRE,Santos2020}, $J_{qcb}$ depends on both $L$ and $S$, so it is difficult to make further analytical considerations concerning the chaos border in both the semiclassical and thermodynamic limit. The results in~\cite{Santos2010PRE,Torres2014PRE,Santos2020} indicate that the quantum chaos border decreases on increasing $L$, as indeed seen in Fig.~\ref{fig:voff}~(b).

Even if each two-body spin Hamiltonian has its particularities, our analysis can be extended to more general Hamiltonians. Instead of our Hamiltonian, we could have a system with additional transverse magnetic fields along the $x,y$-directions that would also couple the nearest-neighbour blocks, $\mu_z \to \mu_z \pm 1$, or it could be a model with only $\hat{V}_{in}$, or other variations.
But overall, the results would be very similar, we would still have very sparse banded matrices due to the presence of two-body couplings. We may therefore say that the Hamiltonian structure analyzed in this work is quite generic for spin systems with two-body interactions. 

\subsection{Level Statistics}

Quantum systems that are chaotic in the classical limit often exhibit correlated eigenvalues as in random matrices~\cite{Guhr1998}. The degree of short-range correlations can be evaluated with the distribution of the spacing $s$ between neighboring unfolded levels. In the case of random matrices belonging to the  Gaussian orthogonal ensemble (GOE), the distribution follows the Wigner surmise, $P_{\text{GOE}}(s) = (\pi s/2) \exp(-\pi s^2/4)$, while uncorrelated levels result in
a distribution close to 
the Poissonian distribution $P_{\text{P}}(s) =  \exp(-s)$ \cite{HaakeBook}.  

A more complete picture of the spectrum requires also the analysis of the long-range correlations. To  measure the rigidity of the spectrum, one resorts to quantities such as the level number variance, $\Sigma^2(\ell)$, which is the variance of the number of unfolded eigenvalues in an interval $\ell$. For GOE, we have that $\Sigma^2_{\text{GOE}}(\ell)=2 [\ln(2 \pi \ell) + \gamma + 1 - \pi^2/8]/\pi^2$, where $\gamma$ is the Euler constant, while for the Poissonian case, the fluctuations are larger and the variance grows linearly with the energy interval, $\Sigma^2_{\text{P}}(\ell)=\ell$.

%%%%%%%%%%%%%%%% FIG %%%%%%%%%%%%%%%%%
\begin{figure}[h]
    \centering
    \includegraphics[width=0.45\textwidth]{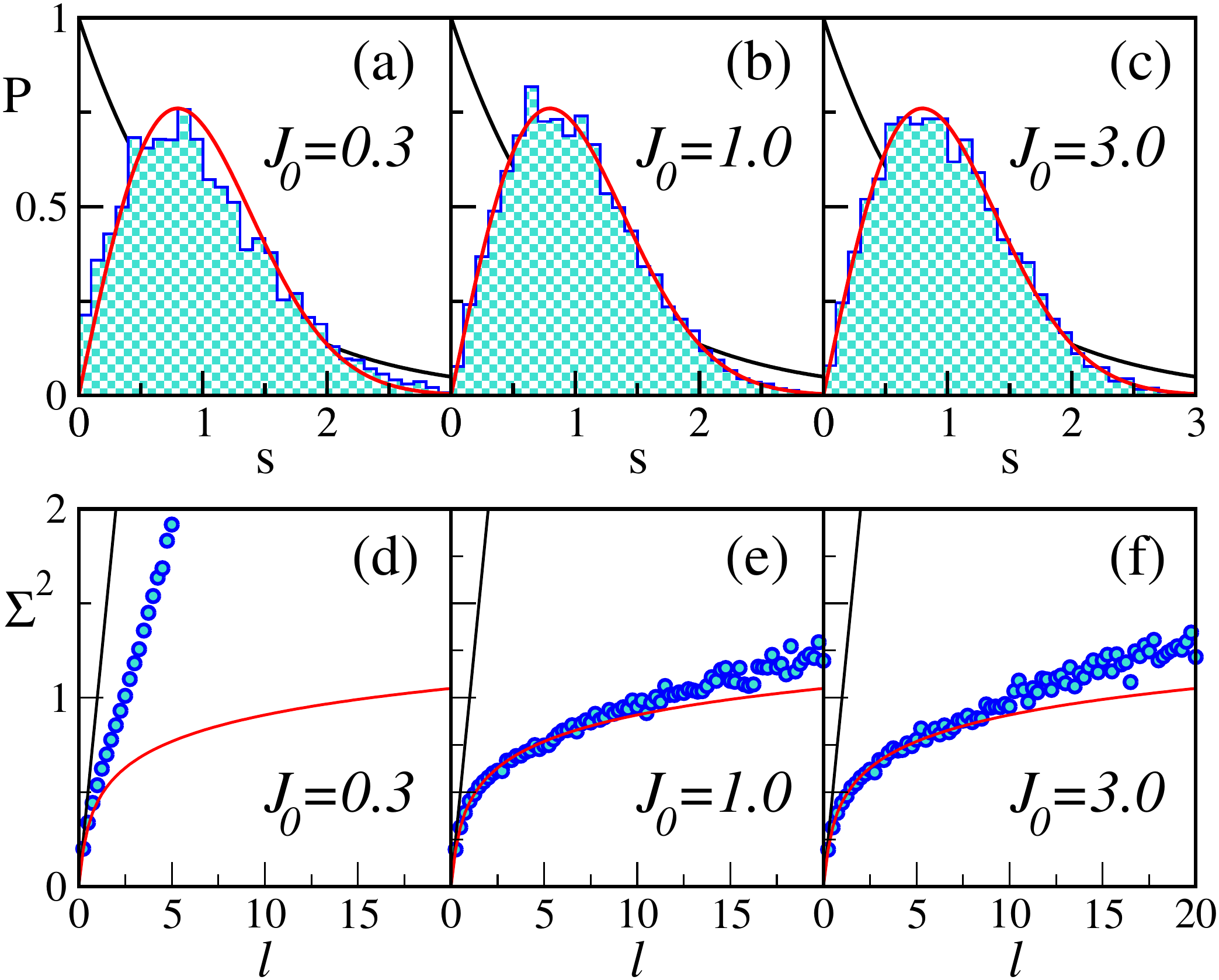}
    \caption{Nearest-neighbor level spacing distribution (a)-(c) and level number variance (d)-(f) for three different interaction strengths; $L=6$, $S=2$, $B_0$=1, $\delta W=0.2$, $\nu=1.4$.   For comparison, the solid lines represent Poisson (black) and GOE (red) results. All the eigenvalues have been computed  from one symmetry sector (even total $z-$magnetization)  discarding 10\% of the levels at the borders.
}
    \label{fig:pofs}
\end{figure}
%%%%%%%%%%%%%%%%%%%%%%%%%%%%%%%%%%

In Fig.~\ref{fig:pofs}, we compare the level spacing distributions [Figs.~\ref{fig:pofs}.~(a)-(c)] and the level number variances [Figs.~\ref{fig:pofs}.~(d)-(f)] for different values of the interaction strength $J_0$, growing from left to right, and keeping fixed the other parameters of the Hamiltonian. 
For all the chosen values of the interaction, the classical system is always chaotic, but deviations from the Wigner-Dyson distribution are seen in Fig.~\ref{fig:pofs}.~(a), because $J_0=0.3$ is below the quantum chaos border. 
Actually, even in the region where good agreement of the level spacing distribution with the Wigner-Dyson distribution is seen [Figs.~\ref{fig:pofs}.~(b)-(c)], deviations from the random matrix theory results appear for the more sensitive measures, such as the level number variance in Figs.~\ref{fig:pofs}.~(e)-(f) for $\ell > 7$. Even though this observation is important, it is not surprising, because we are comparing the spectrum of a sparse banded matrix, that has correlated elements, with the spectrum of a GOE random matrix.

\section{Shape of Eigenfunctions and Local Density of States}
\label{Sec:LDOS}

In this section, we discuss two basic concepts for the study of quantum chaos and show that they have classical analogues. The first one is the envelope of the exact eigenstates written as a function of the non-interacting energies, which is referred to as the ``shape of the eigenstates'' (SoE), and the other is the ``local density of states'' (LDoS).
Both quantities have been extensively investigated  in view of the definition of chaotic eigenstates~\cite{Borgonovi1998,Benet2000}.  Chaotic eigenstates are defined as  eigenstates composed of many components in the non-interacting basis that can be treated as completely uncorrelated.

In physical systems, the eigenstates cannot be fully ergodic in the non-interacting basis due to the finite range of the interactions, which gets reflected in the band-like structure of the total Hamiltonian. Therefore, one can speak at most about pseudo-randomness of the eigenstates in connection with some envelope around which the fluctuations of the squared components of the eigenfunction are Gaussian.
This envelope  can be obtained either by averaging over close eigenstates or by using  moving windows within one  eigenstate. The equivalence of these two averaging procedures is at the core of statistical mechanics. 

%%%%%%%%%%%%%%% FIG %%%%%%%%%%%%%%%%%%% 
\begin{figure*}[ht!]
    \centering
    \includegraphics[width=0.8\textwidth]{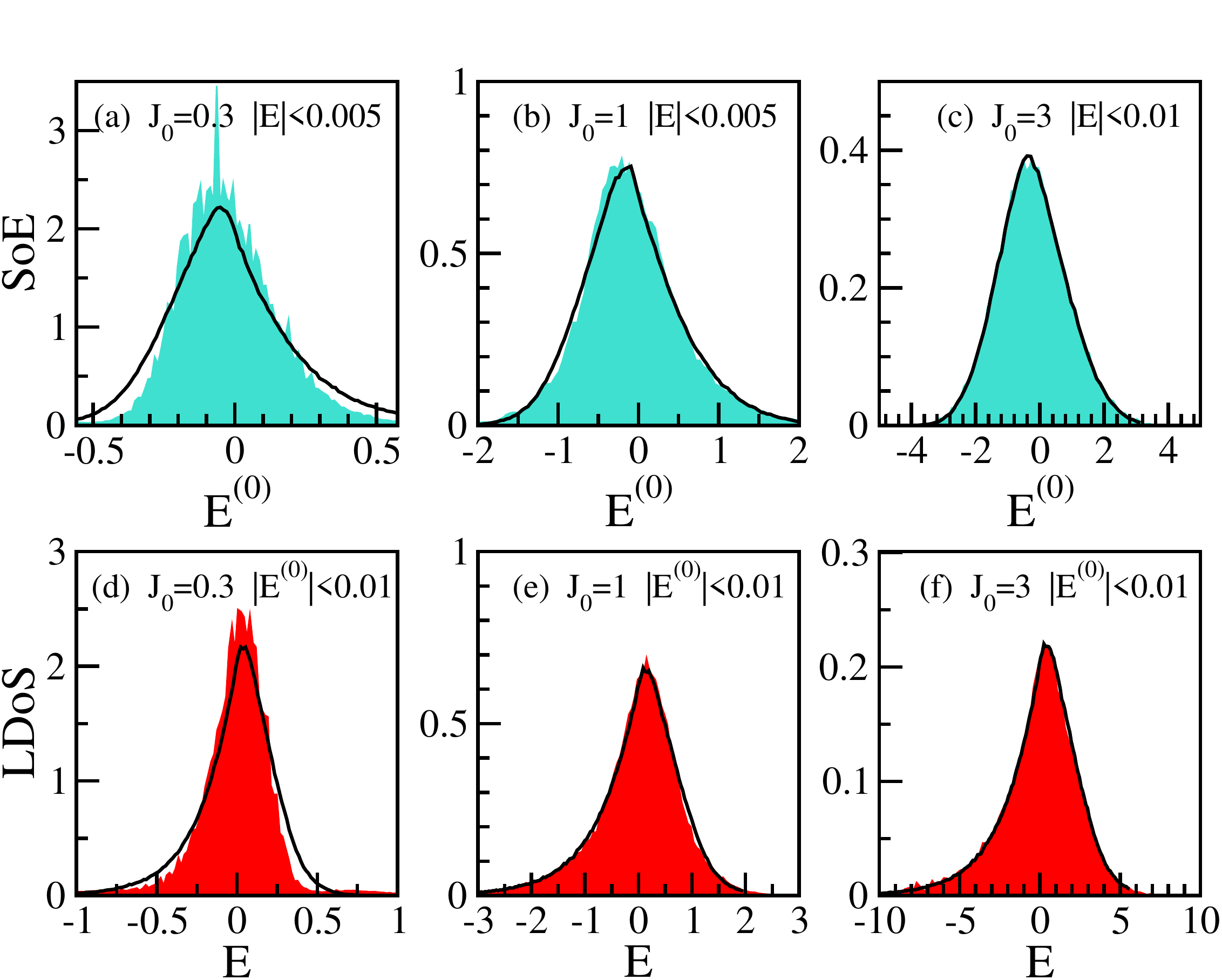}
    \caption{(a)-(c): Quantum (shade) and classical (line) shape of the eigenfunctions; (d)-(f): Quantum (shade) and classical (line)  local density of states. Parameters: $L=6$, $B_0=1$, $\delta W=0.2$, $\nu =1.4$. 
     The quantum-classical correspondence is very accurate  for strong enough interaction, $J_0  \gtrsim 1$.
    For the quantum case, we have $S=2$ and  average over $41$ eigenstates for the SoE and $50$  noninteracting basis states for the LDoS. For the classical functions, we consider a set of $10^5$ initial conditions with total energies (SoE) or noninteracting energies (LDoS)
    in the same interval of the corresponding quantum model. 
}
    \label{fig:sh-ldos}
\end{figure*}
%%%%%%%%%%%%%%%%%%%%%%%%%%%%%%%%%%%%%% 

Both quantities, SoE and LDoS, are broadly employed in the quantum realm. To fix the notation, the Schr\"odinger equation,
\begin{equation}
    \label{eq:S0}
    \hat{H_0}\ket{n} = E^{(0)}_n \ket{n},
\end{equation}
defines the non-interacting eigenstates
$|n\rangle$ and  eigenvalues $E^{(0)}_n $ of $\hat{H}_0$, and
\begin{equation}
    \label{eq:S}
    \hat{H}\ket{\alpha} = E_\alpha \ket{\alpha},
\end{equation}
gives the exact eigenstates $|\alpha\rangle$ and exact eigenvalues $E_\alpha $ for  $\hat{H}$. We use Latin letters for the non-interacting eigenstates and eigenvalues  and Greek for the exact 
ones.
The coefficients 
\begin{equation}
\label{eq:coef}
 C_n^\alpha = \langle n | \alpha \rangle   
\end{equation}
 of the eigenfunctions $\ket{\alpha} = \sum_n C_n^\alpha \ket{n}$ written in the non-interacting basis  are the building blocks of our approach.
They are obtained from  the projection of the exact eigenstates on the non-interacting states or from the projection of the non-interacting states on the exact ones.

\subsection{Shape of the Eigenstates}
For the quantum SoE, we study the components $C_n^\alpha$ as a function of the non-interacting energies $E^{(0)}_n$. We perform an average $\langle .. \rangle$ over the eigenstates in a small energy window, $E-\delta E <E_\alpha <E+\delta E $, and smooth the function,
\begin{equation}
    \label{eq:SOE}
    W_E(E^{(0)}) = \sum_n \delta(E^{(0)}-E^{(0)}_n) \langle |C_n^\alpha |^2 \rangle.
\end{equation}
In this equation the SoE, $W_E(E^{(0)}_n)$, represents the probability that an eigenstate having an energy in the window $[E-\delta E,E+\delta E]$ is found in the non-interacting state of energy $E^{(0)}_n$. 

We obtain the classical analogue of the SoE by taking random initial conditions for spins with energy fixed in a small window  $E-\delta E < H < E +\delta E $ and computing  the non-interacting energy $E^{(0)}_n=H_0( \vec{S}_1,\ldots,\vec{S}_L )$ for all of them. From that, the classical probability $w_E(E^{(0)}_n)$ to have that particular non-interacting energy  (histogram of frequencies) can be obtained.

\subsection{Local Density of States}
As for the quantum LDoS associated with some non-interacting state $\ket{n}$, a similar procedure is used. The LDoS is obtained from   the coefficients $C_n^\alpha$ as a function of the energy of the total Hamiltonian. We perform an average over the non-interacting states in a small energy window,
    $E^{(0)}-\delta E <E^{(0)}_n <E^{(0)}+\delta E $, and smooth the function,
\begin{equation}
    \label{eq:LDOS}
    W_{E^{(0)}}(E) = \sum_\alpha \delta(E-E_\alpha)  \langle |C_n^\alpha |^2 \rangle.
\end{equation}
 This distribution represents the probability for a non-interacting state with energy in the range $[E^{(0)} - \delta E, E^{(0)} + \delta E]$ to be found in the eigenstate with energy  $E$.  
 
Since the quantum LDoS is the energy distribution of a given initial state $|\Psi(0)\rangle $ of the system, it is tightly connected with the system's evolution. The absolute square of the Fourier transform of the LDoS, 
\[
\int W_{E^{(0)}} (E) e^{-i E t/\hbar } dE =|\langle \Psi(0) | \Psi(t) \rangle |^2 ,
\]
for example, is the survival (return) probability, $|\langle \Psi(0) | \Psi(t) \rangle |^2$ , extensively analyzed in studies of quench dynamics (see  \cite{Khalfin1958,Fleming1973,Flambaum2000A,Flambaum2001a,Borgonovi2016,Torres2016Entropy,Schiulaz2019} and references therein), and the width $\sigma_{\text{LDoS}}$ of the LDoS is associated with the lifetime of the initial state. In fact, in the region of strong chaos, $\sigma_{\text{LDoS}}$ is a key parameter for the description of the relaxation process of quantum systems toward thermalization~\cite{Borgonovi2019}.

To construct the classical LDoS, we  fix the non-interacting energy $E^{(0)}$ and take random initial conditions for  each single spin  on the unit sphere with the non-interacting energy $E^{(0)}$   in the required interval  $[E^{(0)}-\delta E, E^{(0)}+\delta E]$. We then compute the total energy  $E$ for all of them and  obtain the probability $w_{E^{(0)}}(E)$ to have the energy $E$  (histogram of frequencies).

\subsection{Quantum-Classical Correspondence} 
The comparison between the classical and quantum SoE and LDoS is done using   the components $C_n^\alpha$ of the eigenfunctions  and the properties of the classical energy surfaces $H_0 = E_0$ and $H=E$. More precisely, the classical distributions can be thought of as a projection of one of these surfaces onto the other~\cite{Casati1993,Casati1996}.
   
 Figures~\ref{fig:sh-ldos}~(a)-(c) show a comparison between the quantum SoE (shade), $W_E(E^{(0)})$, and its classical analogue (solid line), $w_E(E^{(0)})$.
 Figures~\ref{fig:sh-ldos}~(d)-(f) display the comparison between the quantum LDoS (shade),
 $W_{E^{(0)}}(E)$, and its classical version (solid line), $w_{E^{(0)}} (E)$.
There is excellent quantum-classical agreement above the quantum chaos border, $J_0 > 0.5$,  while deviations are seen below the quantum chaos border.  To explain the reason for the discrepancy, it is useful to resort to a  method to compute the classical SoE and LDoS  based on the dynamical equations of motion, as described next.
 
\begin{figure}[ht]
    \centering
    \includegraphics[width=0.45\textwidth]{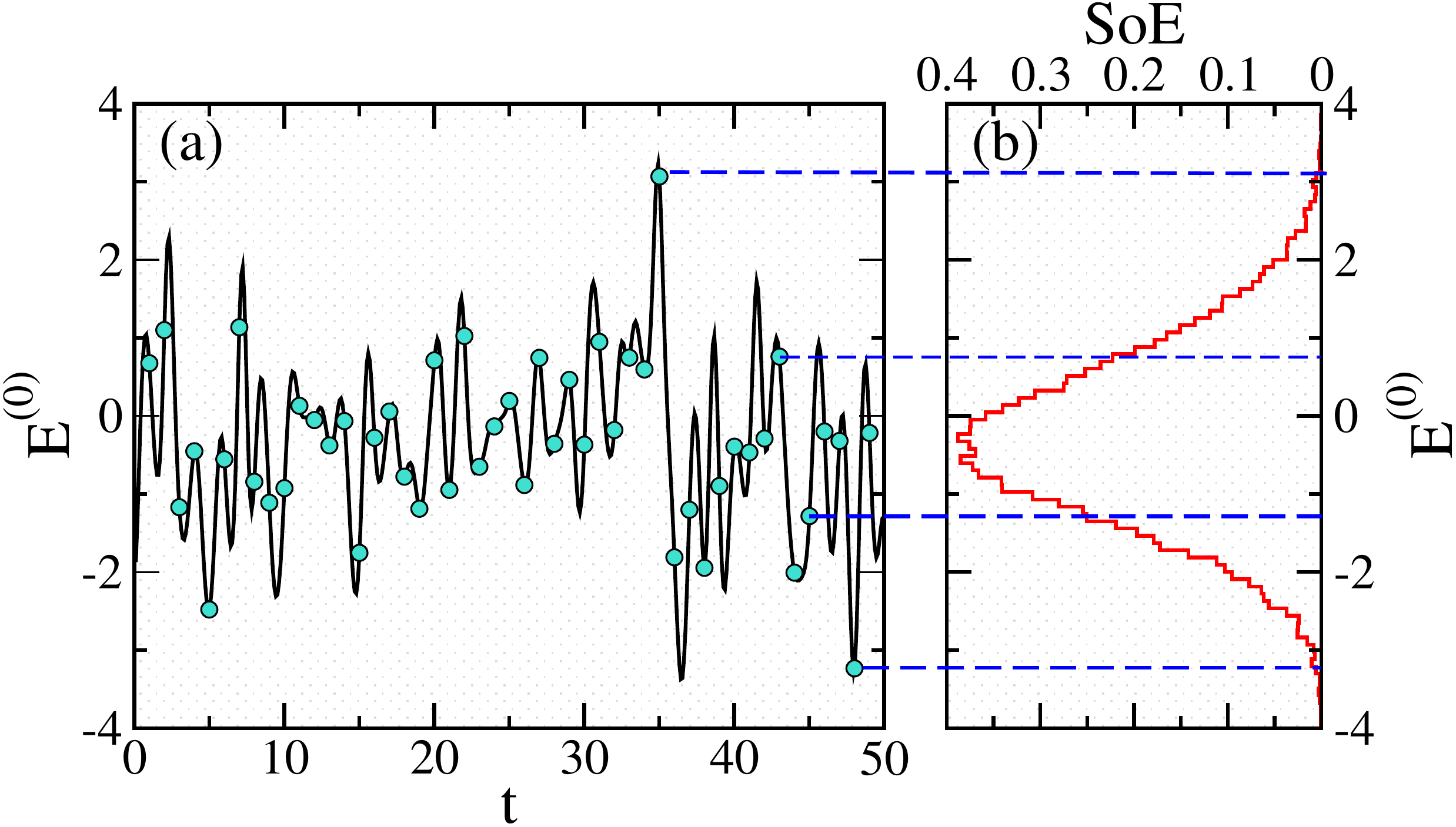}
    \caption{Dynamical construction of the classical shape of the eigenfunctions. In (a): Trajectory $E^{(0)}(t)$ obtained by inserting in the noninteracting Hamiltonian $H_0$ the solution of the full equations of motion under the total Hamiltonian $H$. Random sample of values of $E^{(0)}(t)$ at equally spaced times are represented by circles. In (b): These points, 
    which represent how a trajectory at some specified energy fills the ``energy shell'' of  $H_0$, are used  to plot the normalized probability distribution function.
}
    \label{fig:dyn-h}
\end{figure}

The procedure to obtain SoE goes as follows.  We  consider   one initial condition 
$u(0) = \vec{S}_1(0),\ldots,\vec{S}_L (0)$
with energy $E$ in the chosen energy window and consider its evolution $u(t)$ under the full interacting Hamiltonian $H$. From that we  compute the function $E^{(0)}(t) = H_0(u(t)) = H_0(\vec{S}_1(t),\ldots,\vec{S}_L (t))$ at several equally spaced times $t_k = k\Delta t$, where $\Delta t$ is chosen of the same order of the typical period of the motion, $1/B_0$, to have statistical independent data. The function $E^{(0)}(t)$ is shown in Fig.~\ref{fig:dyn-h}~(a) and the values at $t_k$ are indicated with circles. The normalized distribution of these points is the SoE shown in Fig.~\ref{fig:dyn-h}~(b). 
%%%%%%%%%%%%%%% FIG %%%%%%%%%%%%%%%%%%% 
%%%%%%%%%%%%%%%%%%%%%%%%%%%%%%%%%%%%%%%
It gives the same result as the ``static'' method used to obtain Fig.~\ref{fig:sh-ldos}~(a) provided the trajectory can cover ergodically  the whole energy range obtained by the phase space sampling. Most importantly, it is only in this situation that we can properly define the classical SoE. In the case in which, due to some physical or dynamical constraints,  different trajectories produce different distribution functions, then a proper definition of the classical SoE becomes problematic. We then have the following non-trivial result:  {\it the quantum SoE admits a well-defined classical limit only if the classical motion is ergodic in some energy region. }

This analysis can be equivalently extended to the LDoS, where the dynamical and static methods to compute the classical LDoS give equal results only in the region of strong quantum chaos, and they also coincide with the quantum LDoS. Analogously to SoE, the dynamical construction of the classical LDoS requires a single  initial condition 
$u_0(0) = \vec{S}_1(0),\ldots,\vec{S}_L (0)$
with fixed non-interacting energy $E^{(0)}$ and its evolution $u_0(t)$ under the non-interacting Hamiltonian $H_0$.  

To compute $u_0(t)$ under the non-interacting Hamiltonian we do not need to numerically integrate the equations of motion. The classical evolution under $H_0$ corresponds simply to rotations  of all spins about the $z$-axis with frequencies $B_k$, and it is given by,
\begin{equation}
    \label{eq:nonint}
    \begin{array}{lll}
         (S_k^x)^0(t) &= S_k^x(0) \cos(B_k t) + S_k^y(0) \sin(B_k t)   \\
         & \\
         (S_k^y)^0(t) &= S_k^y(0) \cos(B_k t) - S_k^x(0) \sin(B_k t)   \\
         & \\
         (S_k^z)^0(t) &= S_k^z(0).
    \end{array}
\end{equation}
With the trajectory under the non-interacting Hamiltonian $H_0$, we obtain the 
  function $E(t) = H(\vec{S}_1^0(t),\ldots,\vec{S}_L^0 (t))$ from where we extract the values at $t_k = k\Delta t$, with $ \Delta t \sim 1/B_0$, to build the correspondent  normalized histogram.
  
\begin{figure}[h]
    \centering
    \includegraphics[width=8.5cm]{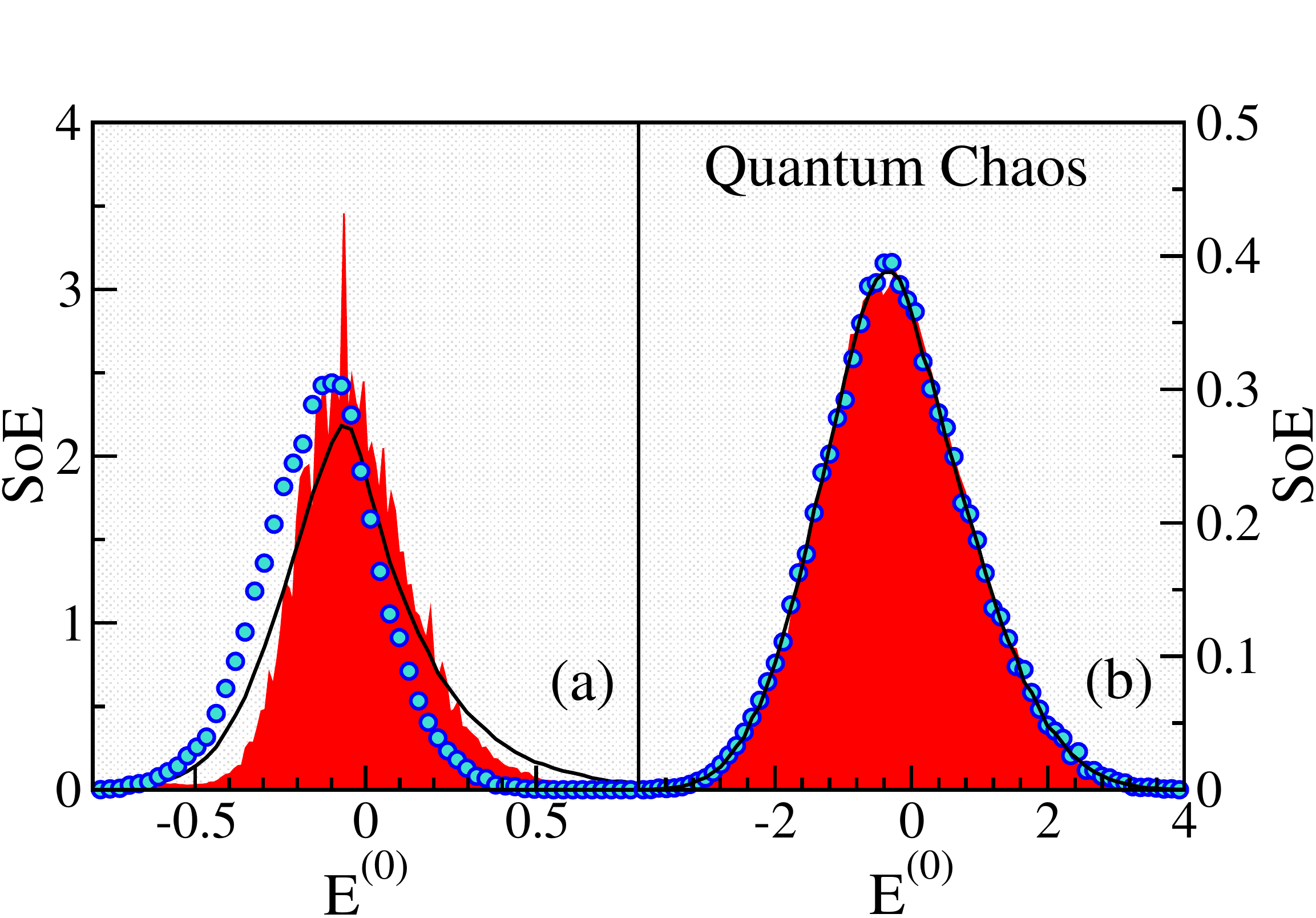}
    \caption{Classical and quantum shape of the eigenfunctions for two different interaction strengths: $J_0=0.3$ (a)
    and $J_0=3$ (b). 
    (Red) histograms represent the quantum data obtained by averaging 41 eigenfunctions in a small energy window in the centre of the spectrum. (Black) full lines stand for the classical SoE obtained by averaging over $10^5$ random points in the same small energy window (static method). (Blue) circles are  obtained with a single trajectory with energy $|E|<0.01$ (dynamical method). 
}
    \label{fig:erg}
\end{figure}

To understand the mechanism for the discrepancy  between the classical and quantum quantities  for small interaction strength, $J_0<0.5$, as  shown in Figs.~\ref{fig:sh-ldos}~(a),(d), we take the case of the SoE as an example. Needless to say a similar analysis can be done for the LDoS.
In Fig.~\ref{fig:erg} we compare the quantum SoE (shade area) with the classical SoE obtained with the static method (line) and with the dynamical method (circles) for two values of the interaction strength. While the three distributions coincide in the quantum chaotic regime  [Fig.~\ref{fig:erg}~(b)], they differ for small interaction strength [Fig.~\ref{fig:erg}~(a)].  The lack of agreement between the two classical distributions in Fig.~\ref{fig:erg}~(a) is a clear indication of the lack of ergodicity. Due to dynamical reasons, such as the presence of islands of stability or dynamical constraints, a single trajectory cannot span the whole allowable energy range as defined by the random initial conditions used to implement the static distribution. Surprisingly, even the quantum distribution differs from the classical static one. The fact that the quantum distribution is narrower suggests  the presence of quantum localization. This should not to be confused with the lack of classical chaos, since both cases, $J_0=0.3$ and $J_0=3$, are characterized by the exponential divergence of close trajectories, signalled by a positive maximal Lyapunov exponent.
%%%%%%%%%%%%%. FIGURE%%%%%%%%%%%%%%%%%%%%%%%%%%%%%%%%%%%%
%%%%%%%%%%%%%%%%%%%%%%%%%%%%%%%%%%%%%%%%%%%%%%%%%%%%%%

\section{Kolmorogov-Sinai entropy and width of the LDoS}
\label{Sec:KS-LDOS}

As presented in Fig.~2, the rate of the exponentially growing separation between close trajectories for each individual spin is related with the maximal exponent of the full Lyapunov spectrum of the whole spin system. But what is the role of the other positive Lyapunov exponents?
To address this question, we resort to the studies in Zaslavsky's book~\cite{ZaslavskyBook}, where he compares the value of the Kolmogorov-Sinai entropy with the exponential growth in time of the coarsed grained phase-space volume for  multidimensional classical systems with chaotic behavior, as in our  Eq.~(\ref{eq:Zy}). We have discussed this relation in our previous work~\cite{Borgonovi2019} about the onset of quantum chaos in fermionic and bosonic systems  characterized by two-body interactions.  Specifically, we have argued that the Kolmogorov-Sinai entropy [$h_{\text{KS}}$ in Eq.~(\ref{eq:hks})]  should be directly related to the  width of LDoS [$\sigma_{\text{LDoS}}$ in Eq.~(\ref{eq:LDOS})]. We now confirm this expectation for our spin system.

%%%%%%%%%%%%%% FIG %%%%%%%%%%%%%%%
\begin{figure}
    \centering
    \includegraphics[width=0.45\textwidth]{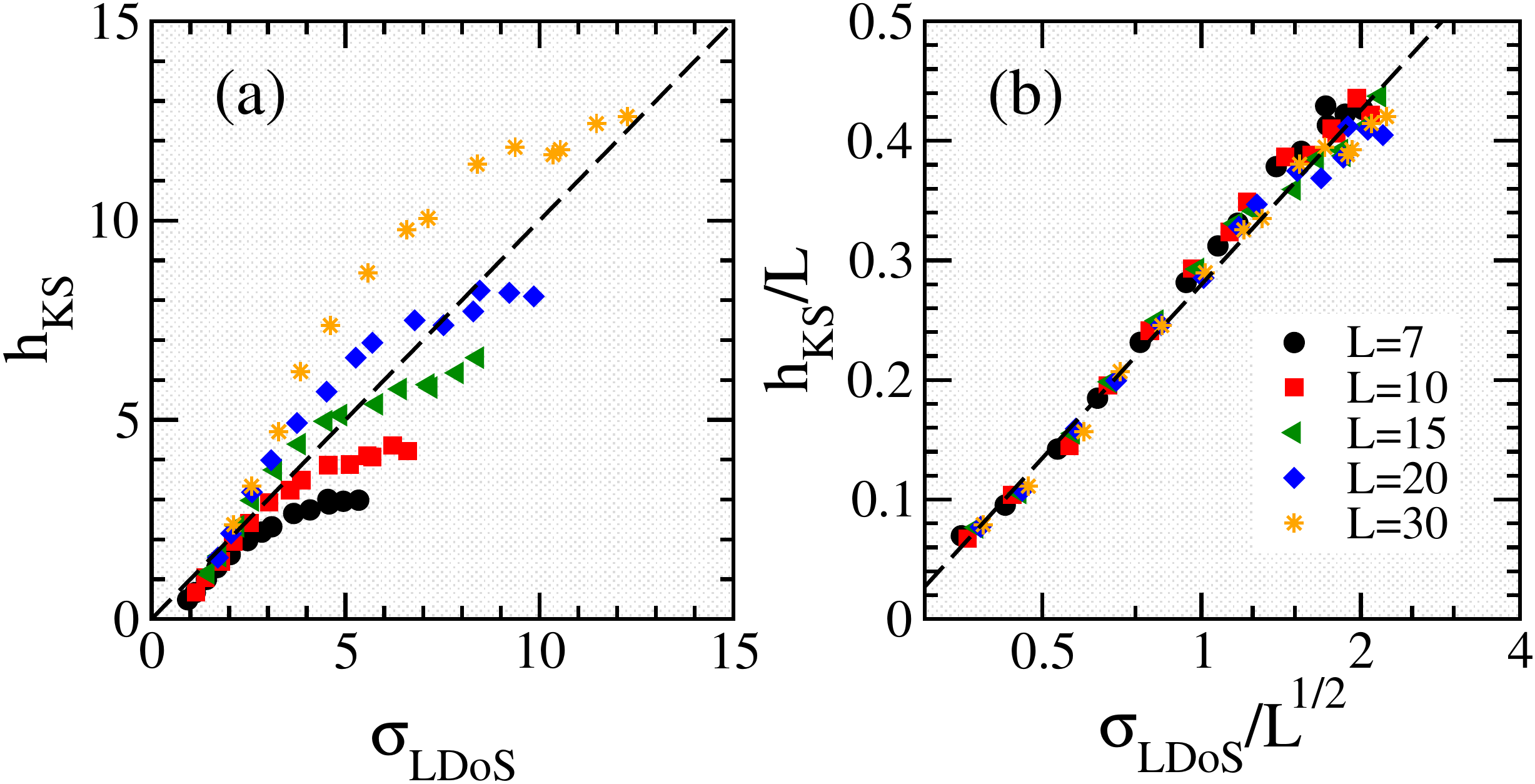}
    \caption{  (a) Comparison between the Kolmogorov-Sinai entropy and the width of the LDoS for 
  $B_0=1$, $W=0.2$, $\nu =1.4$, initial  energy $|E|<0.01$, and different interaction strengths $0.5<J_0<10$.
  The dashed line  guides the eye and represents the line $y=x$.
  (b) Comparison between the density of the Kolmogorov-Sinai entropy and the rescaled width of the LDoS for the same data as in panel (a). In the $x$-axis the logarithmic scale has been used.  Here, the dashed line stands for the best logarithmic fit  $y=0.28+0.21 \ln (x)$.
}
    \label{fig:s1}
\end{figure} 
%%%%%%%%%%%%%%%%%%%%%%%%%%%%%%%%%%%

Figure~\ref{fig:s1}~(a) compares $\sigma_{\text{LDoS}}$ and $h_{\text{KS}}$  in the energy range of maximal chaos ($|E|<0.01$) for different system sizes $L$ and interaction strengths $J_0$. Different colors indicate different system sizes, and the interaction strength  for each set of data grows from the left to the right in the panels.
In the chaotic region considered, which is characterized  by ergodicity, the classical and quantum LDoS are very close, so one can also see the comparison between the Kolmogorov-Sinai entropy and the width of LDoS in Fig.~\ref{fig:s1} as  a comparison  between two well defined, but physically different classical quantities. 

In Fig.~\ref{fig:s1}~(a), we observe that $h_{\text{KS}}$ and   $ \sigma_\text{LDoS}$
  are on the same order of magnitude,  but deviations are visible, which depend on both the values of  $L$ and $J_0$.  In Fig.~\ref{fig:s1}~(b), we show the same data, but rescaled as $\Tilde{h}_{\text{KS}}=h_{\text{KS}}/L$ (density of the Kolmogorov-Sinai entropy) and $\Tilde{\sigma}_{\text{LDoS}}=\sigma_\text{LDoS}/\sqrt{L}$ (renormalized energy width).
The reason for the rescaling with $\sqrt{L}$ for the width of the LDoS is that in the chaotic regime, the classical width of LDoS can be considered as the sum of $L$ independent  random variables $S_k^z$, whose second moment is proportional to  $L$. With this rescaling,  all points collapse onto a single curve well described by a  logarithmic fit.  

The numerically found logarithmic dependence, 
\begin{equation}
\Tilde{h}_{\text{KS}}  \propto \ln \Tilde{\sigma}_{\text{LDoS}},
\end{equation}
is  a relationship between two intensive quantities. 
This remarkable result should be checked  in other  models as well. This is a very important finding, because
these two  quantities have a completely different dynamical origin. The Kolmogorov-Sinai entropy is directly related to the {\it local} instability of close trajectories, while  the LDoS  is associated with the 
{\it global } properties of the relaxation process. 
We hope that the relation above will trigger future investigations in the field.

%%%%%%%%%%%%%%%%%%%%%%%% CONCLUSION %%%%%%%%%%%%%%%%%%%%%%
\section{Summary}
\label{sec:conclusions}
 
The aim of our study was to establish the quantum-classical correspondence (QCC) for interacting spin-models, which can be strongly chaotic in the classical limit. Our results indicate that this correspondence holds only  in the region of strong classical and quantum chaos, which corresponds to the energy region  $E \approx 0$, as confirmed from the analysis of the Lyapunov spectrum.
 
Starting with the analysis of the classical equations of motion, we found that in addition to energy, our model presents other integrals of motion. This property, which is generic to spin systems, stems from the fact that the motion of each individual spin is restricted to a 3D sphere, which results in a many-dimensional phase space with a non-standard Hamiltonian structure. 

We observed that for weak interaction, the motion of each individual spin can be effectively described as a {\it linear} parametric oscillator under an external force consisting of a large number of  harmonics. In this case, one can speak of the emergence of {\it linear chaos}, a term introduced by Chirikov~\cite{ChirikovProceed,Chirikov1997}.

As the interaction strength increases,  the influence of non-linear resonances emerge due to the non-linear coupling between the spins. However, our results made it clear that in many aspects the dynamical properties of the model can still be effectively described with the motion of individual spins. In particular, we showed that ergodicity of the full model boils down to ergodicity of the motion of each individual spin. This allows for the introduction of a test of the {\it local ergodicity} of the motion of individual spins. Following random matrix theory, ergodic motion of the trajectory on a sphere implies a flat distribution of the components of each spin. Our numerical data confirmed the emergence of this flat distribution for a large enough interaction strength ($J_0 \gtrsim 1$) and sufficiently long spin chains ($L \gg 1$). One can therefore speak of {\it global ergodicity}, when the motion of {\it all} spins are {\it locally ergodic}.

Our study of the quantum counterpart of the system started with the analysis of the structure of the Hamiltonian matrix, which allowed us to establish a criterion for the onset of quantum chaos. We then showed that the shape of the eigenfunctions (SoE) and the local density of states (LDoS), which are essential quantum quantities, have classical analogues that can be obtained using the classical equations of motion. There is an excellent correspondence between the classical and quantum distributions in the region of strong quantum chaos, but not when ergodicity is broken.

In the last section, we presented a compelling relationship between the Kolmogorov-Sinai entropy, $h_{\text{KS}}$, and the width of the LDoS, $\sigma_{\text{LDoS}}$. Since $h_{\text{KS}}$ is related to the local instability of motion, it is not a directly measurable quantity, but $\sigma_{\text{LDoS}}$ is the decay rate of the survival (return) probability, which is a global quantity that can be measured after a quantum quench. This opens the possibility to relate the Kolmogorov-Sinai entropy with a physically measurable quantity.

%%%%%%%%%%%%%%%%%%%%%%%%%%%%%%%%%%%%%%%%%%%%%%%%%%%%%%
\begin{acknowledgments}
 F.B.  acknowledges support by the Iniziativa Specifica INFN-DynSysMath. This work
has been financially supported by the Catholic University of Sacred Heart within the program of promotion and diffusion of scientific research. Research has also been financially supported by Ministero dell’Istruzione, dell’Universit\'a e della Ricerca within the Project No. PRIN 20172H2SC4.
F.M.I. acknowledges financial support from CONACyT (Grant No. 286633). L.F.S. had support from the MPS Simons Foundation Award ID: 678586.
This research was supported by the United States National Science Foundation (NSF) Grant No. DMR-1936006. L.B. acknowledges support from UNAM--PAPIIT grant number IG-101122.
\end{acknowledgments}

%%%%%%%%%%%%%%%%%%%%%%%%%%%%%%%%%%%%%%%%%%%%%%%%%%%%%%
%\bibliography{biblioNSF2019}
%%%%%%%%%%%%%%%%%%%%%%%%%%%%%%%%%%%%%%%%%%%%%%%%%%%%%%

%apsrev4-2.bst 2019-01-14 (MD) hand-edited version of apsrev4-1.bst
%Control: key (0)
%Control: author (8) initials jnrlst
%Control: editor formatted (1) identically to author
%Control: production of article title (0) allowed
%Control: page (0) single
%Control: year (1) truncated
%Control: production of eprint (0) enabled
%

\end{document}